\documentclass[12pt]{article}
\usepackage[english]{babel}
\usepackage[utf8]{inputenc}
\usepackage{graphicx}
\usepackage{cancel}
\usepackage{amsmath,amssymb,amsfonts,amsthm}
\usepackage{geometry}
\usepackage{mathrsfs}
\usepackage{nicefrac}
\usepackage{booktabs}
\usepackage{multirow}
\usepackage{enumitem}
\usepackage{float}
\usepackage{hyperref}
\usepackage{blkarray}
\hypersetup{colorlinks = true, linkcolor=red, citecolor=blue}
\usepackage{mathtools}
\usepackage{upgreek}

\usepackage[ruled]{algorithm2e}
\usepackage{authblk}
\usepackage{subcaption}
\usepackage{bm} 
\usepackage{todonotes}
\DeclareUnicodeCharacter{202F}{FIX ME!!!!}

\newcommand{\R}{\mathbb R}
\newcommand{\N}{\mathbb N}

\newcommand{\C}{\mathbb C}

\newcommand\norm[1]{\lVert #1 \rVert}

\newcommand\inner[1]{\langle #1 \rangle}

\DeclareMathOperator{\rank}{rank}

\DeclareMathOperator{\ran}{ran}

\newcommand{\ra}{\rightarrow}

\numberwithin{equation}{section}
\newtheorem{theorem}{Theorem}[section]
\newtheorem{proposition}[theorem]{Proposition}
\newtheorem{lemma}[theorem]{Lemma}
\newtheorem{corollary}[theorem]{Corollary}

\theoremstyle{definition}
\newtheorem{definition}[theorem]{Definition}
\newtheorem*{definition*}{Definition}
\newtheorem{assumption}{Assumption}

\newtheorem*{remark*}{Remark}

\begin{document}

\title{A mathematical analysis of the adiabatic Dyson equation from time-dependent density functional theory
}

\author{Thiago Carvalho Corso\thanks{Email: \url{thiago.carvalho@ma.tum.de}}}
\affil{Zentrum Mathematik, Technische Universit\"at M\"unchen, Germany}

\renewcommand\Affilfont{\itshape\small}

\maketitle

\begin{abstract}

In this article, we analyze the Dyson equation for the density-density response function (DDRF) that plays a central role in linear response time-dependent density functional theory (LR-TDDFT). First, we present a functional analytic setting that allows for a unified treatment of the Dyson equation with general adiabatic approximations for discrete (finite and infinite) and continuum systems. In this setting, we derive a representation formula for the solution of the Dyson equation in terms of an operator version of the Casida matrix. While the Casida matrix is well-known in the physics literature, its general formulation as an (unbounded) operator in the $N$-body wavefunction space appears to be new. Moreover, we derive several consequences of the solution formula obtained here; in particular, we discuss the stability of the solution and characterize the maximal meromorphic extension of its Fourier transform. We then show that for adiabatic approximations satisfying a suitable compactness condition, the maximal domains of meromorphic continuation of the initial density-density response function and the solution of the Dyson equation are the same. The results derived here apply to widely used adiabatic approximations such as (but not limited to) the random phase approximation (RPA) and the adiabatic local density approximation (ALDA). In particular, these results show that neither of these approximations can shift the ionization threshold of the Kohn-Sham system.

\end{abstract}

\tableofcontents

\section{Introduction}

Time-dependent density functional theory (TDDFT) is a formally exact theory to study the time evolution of a system of electrons; it has many applications in quantum chemistry, condensed-matter physics, and material science \cite{burkeDFTreview,CASIDA2009,ullrich2012time,Vasiliev2002}. Most of these applications lie within the perturbative regime, where linear response theory applies (LR-TDDFT). In this regime, one is no longer interested in the whole non-linear evolution of the single-particle density of the system but instead in the linear dynamical response of the density to a variation of the external potential. Stated differently, one is interested in the \emph{density-density response function} (DDRF) of the system. 

The fundamental equation of LR-TDDFT is the celebrated Dyson equation that formally connects the DDRF of a given system of interest to the DDRF of an equivalent system of non-interacting electrons, the Kohn-Sham system. The equivalence is in the sense that both systems have the same ground state density. In shorthand notation, the Dyson equation reads\footnote{For the precise definition of the product $\chi_0 \star F_{\rm Hxc} \chi$ and a formal derivation of the Dyson equation, we refer to \cite[Appendix]{DDRF2023}.}
\begin{align*}
    \chi = \chi_0 + \chi_0 \star F_{\rm Hxc} \chi, 
\end{align*}
where $\star$ denotes the convolution in time, $\chi$ is the DDRF of the system of interest, $\chi_0$ is the DDRF of the Kohn-Sham system, and $F_{\rm Hxc}$ is the linear operator whose Schwartz kernel is the Hartree plus exchange-correlation (Hxc-)kernel of TDDFT. In principle, the \emph{Hxc-operator} depends on the ground state density of the system or, equivalently, on the Kohn-Sham ground state density. In practice, this density dependence is highly non-trivial, and the exact Hxc-operator is unknown; one then relies on approximations of this operator. 

While several classes of approximations for the Hxc-operator were suggested in the physics literature (see \cite{marques2012fundamentals,ullrich2012time} for an overview), the overwhelming majority of calculations are performed with \emph{adiabatic approximations}. In the adiabatic approximation, the Hxc-operator acts instantaneously on the perturbing potential and can be seen as an operator acting on the spatial part of the perturbation. Within the adiabatic approximation, the Dyson equation becomes
\begin{align}
\chi_{F}(t) = \chi_0(t) + \int_0^t \chi_0(t-s) F \chi_{F}(s) \mathrm{d} s, \label{eq:Dysonadiabatic}
\end{align}
where the adiabatic approximation $F$ is now an operator from the (tangent) space of densities to the (tangent) space of potentials, $\chi_0$ is again the DDRF of the Kohn-Sham system, and $\chi_{F}$ is now an approximation of the true DDRF of the system of interest. For suitable choices of $F$, such approximations are observed to reproduce many response properties of large quantum systems accurately (see, e.g., \cite{Vasiliev2002}).

In the non-perturbative regime, TDDFT models were considered in various settings (see, e.g., \cite{ChadamHartree1975,CancesHartree1999,CancesRPA2012,lewin2014hartree,SprengelTDKS2017,pusateri2021long}). By contrast, the mathematical literature on the linear response regime is scarce. Some works \cite{BaiCangLReigenvalueI2012,BaiCangLReigenvalueII2013,Brabec2015,Toulouse2022} have focused on the numerical aspects of extracting relevant properties (e.g., excitation energies, oscillator strengths, and absorption spectrum cross-section) of the solution $\chi_F$ in the finite-dimensional case, i.e., after discretization of the underlying function space. However, a first step towards a rigorous understanding of the Dyson equation \eqref{eq:Dysonadiabatic} in the infinite-dimensional (continuum) setting has been taken only recently in \cite{DDRF2023}. There, the authors presented a mathematical framework for studying the Dyson equation within the Random Phase Approximation (RPA). 

The current paper extends the framework presented in \cite{DDRF2023} to deal with more general adiabatic approximations. Moreover, the main contributions of this article can be summarized as follows:
\begin{enumerate}[label=(\arabic*)]
    \item We generalize the functional analytic setting presented in \cite{DDRF2023} to allow for a unified treatment of the Dyson equation with general adiabatic approximations for both discrete (finite and infinite) and continuum systems. Notably, this new setting allows us to study the celebrated adiabatic local density approximation (ALDA).
    
    \item We derive an explicit representation formula for the solution $\chi_F$ in the case where $\chi_0$ is the DDRF of a self-adjoint Hamiltonian. For this, the key result is a representation formula for the Fourier transform $\widehat{\chi_F}$ in terms of an operator version of the Casida matrix.
    
    \item We derive and discuss several consequences of this representation formula. In particular, we characterize the maximal meromorphic extension of $\widehat{\chi_F}$ and show that, for widely used adiabatic approximations such as the RPA and ALDA, the maximal domain of meromorphic continuation of $\widehat{\chi_F}$ and $\widehat{\chi_0}$ are the same. Physically, this means that these approximations are not able to shift the ionization threshold of the Kohn-Sham system.
\end{enumerate}

\begin{remark*}[Response theory terminology] In the physics literature, the name density-density response function usually refers to the Schwartz kernel of $\chi_H(t)$. Here we refer instead to the operator-valued function $t \mapsto \chi_H(t)$ as the density-density response function. Let us also remark that $\chi_H$ is sometimes called the (linear) susceptibility \cite{burkeDFTreview} or the reducible (or irreducible) polarizability operator \cite{lin2019mathematical}. \end{remark*}

\section{Main results}

We now introduce some notation and discuss our main results. Throughout this article, $H$ is a self-adjoint operator acting on the anti-symmetric $N$-fold tensor product of $L^2(\Omega,  \mathrm{d}\mu)$, 
\begin{align}
	\mathcal{H}_N \coloneqq \bigwedge_{j=1}^N L^2(\Omega, \mathrm{d} \mu), \label{eq:Nbodyspace}
\end{align}
where $(\Omega, \mu)$ is a measure space. The specific measure space is not relevant to our results; in particular, $\mu$ can be the counting measure on some countable set $\Omega \subset \R^n$ (discrete systems), the Lebesgue measure on some open set $\Omega \subset \R^n$ (continuum systems), or a combination of both (continuum systems with internal spin). Moreover, we assume the following.
\begin{assumption} \label{assump:main} The self-adjoint operator $H : D(H)\subset \mathcal{H}_N \rightarrow \mathcal{H}_N$ satisfies the following:
\begin{enumerate}[label=(\roman*)]
\item \label{it:spectralgap}(Spectral gap) The ground state energy $\mathcal{E}_0 \coloneqq \inf \sigma(H) > -\infty$ is a simple eigenvalue. 
\item\label{it:complexconj}(Real Hamiltonian) $H$ commutes with complex conjugation.
\end{enumerate}
\end{assumption}
Since the ground state of $H$ is non-degenerate, we can unambiguously define its ground state single-particle density (or simply density) as
\begin{align}
\rho_0(r) \coloneqq N \int_{\Omega^{N-1}} |\Psi_0(r,r_2,...,r_N)|^2 \mathrm{d}\mu(r_2) ... \mathrm{d} \mu(r_N), \label{eq:groundstatedensity}
\end{align}
where $\Psi_0$ is the unique (up to phase) normalized ground state wave function of $H$. We then introduce the norms
\begin{align}
	\norm{f}_{\rho_0} =  \biggr(\int_{\Omega} |f(r)|^2 \rho_0(r) \mathrm{d} \mu(r)\biggr)^{\frac12}\quad\mbox{and}\quad\norm{f}_{\nicefrac{1}{\rho_0}} = \biggr(\int_{\Omega} |f(r)|^2 \rho_0(r)^{-1} \mathrm{d} \mu(r)\biggr)^{\frac12} ,
\end{align}
and define the respective weighted $L^2$ spaces as
\begin{align}
	&L^2_{\rho_0} = \{ f : \mathrm{supp}(\rho_0) \rightarrow \C \mbox{ $\mu$-measurable }: \norm{f}_{\rho_0} < \infty \} , \\
&L^2_{\nicefrac{1}{\rho_0}} = \{ f: \mathrm{supp}(\rho_0) \rightarrow \C \mbox{ $\mu$-measurable :} \norm{f}_{\nicefrac{1}{\rho_0}} < \infty \},
\end{align}
where supp$(\rho_0)$ denotes the support of $\rho_0$. As usual, we identify the functions that coincide $\mu$-almost everywhere. Let us also introduce the reduced Hamiltonian 
\begin{align}
    H_\# \coloneqq P_{\Psi_0^\perp} \bigr(H - \mathcal{E}_0\bigr) P_{\Psi_0^\perp}, \label{eq:reducedH}
\end{align}
where $P_{\Psi_0^\perp}$ is the orthogonal projection on $\{\Psi_0\}^\perp$. Note that $H_\#$ is a positive operator acting on $\{\Psi_0\}^\perp$ with domain $D(H_\#) = D(H) \cap \{\Psi_0\}^\perp$. 

\subsection{The density-density response function}

The density-density response function (DDRF) of $H$ is the operator-valued function
\begin{align}
    t \in \R &\mapsto \chi_H(t)  = -2\theta(t) B \sin\bigr( tH_\# \bigr) B^\ast,\quad  \label{eq:chidef}
\end{align}
where $\theta(t)$ is the Heaviside step function, $\mathcal{E}_0$ is the ground state energy of $H$, and the operators $B = B_{\Psi_0}$\footnote{Here we use the operator $B$ introduced in \cite{GW0method2016} instead of the operator $S\Phi(r) = N \int_{\Omega^{3N-N}} \overline{\Psi_0}(r,...) \Phi(r,...)$ introduced in \cite{DDRF2023}. This conveniently reduces the notation because $B =  SP_{\Psi_0^\perp}$ and we mostly work in the space $\{\Psi_0\}^\perp$.} and $B^\ast = B^\ast_{\Psi_0}$ are defined as follows:
\begin{align}
   &\bigr(B \Phi\bigr)(r) = N\int_{\Omega^{N-1}} \overline{\Psi_0(r,r_2,...,r_N)} \Phi(r,r_2,...,r_N) \mathrm{d}\mu(r_2) ... \mathrm{d}\mu(r_N) - \inner{\Psi_0,\Phi}_{\mathcal{H}_N} \rho_0(r),\label{eq:Bdef} \\
   &\bigr(B^\ast f\bigr) (r_1,...,r_N) = \biggr(\sum_{j=1}^N f(r_j) - \inner{\rho_0, f}_{L^2(\Omega,d\mu)}\biggr) \Psi_0(r_1,...,r_N) . \label{eq:Badjointdef}
\end{align}
The connection of this definition with the linear response of the system associated with $H$ can be found in \cite{DDRF2023} and will not be discussed here. Instead, we shall limit ourselves to describing the main properties of $\chi_H$ and presenting a characterization of the maximal meromorphic extension of its Fourier transform. 

Let us start by recalling some results from  \cite{DDRF2023}. In \cite{DDRF2023}, it is shown that the DDRF of typical Schr\"odinger operators on $\R^{3N}$ is a uniformly bounded and strongly continuous function with values in the set of bounded operators between suitable $L^p$ spaces. By adapting the proof in \cite[Section 2]{DDRF2023} to the current setting, we can show that $\chi_H$ is, in fact, uniformly bounded and strongly continuous on the space of bounded linear operators from $L^2_{\rho_0}$ to $L^2_{\nicefrac{1}{\rho_0}}$, denoted here by $\mathcal{B}(L^2_{\rho_0},L^2_{\nicefrac{1}{\rho_0}})$. Consequently, the Fourier transform of $\chi_H$, defined via 
\begin{align}
    \widehat{\chi_H}(\omega) = \lim_{\eta\ra 0^+} B\bigr((\omega+i\eta - H_\#)^{-1} - (\omega+i\eta + H_\#)^{-1}\bigr)B^\ast ,\label{eq:chiHFT}
\end{align}
is a tempered distribution with values on $\mathcal{B}(L^2_{\rho_0},L^2_{\nicefrac{1}{\rho_0}})$. 
(See, e.g., \cite[Proposition 2.8]{DDRF2023} for a derivation of \eqref{eq:chiHFT}). In particular, $\widehat{\chi_H}$ has an analytic extension to the upper half-plane. With some abuse of notation, we denote this extension also by $\widehat{\chi_H}$.

An immediate consequence of the spectral gap assumption on $H$ is that $\widehat{\chi_H}$ can be analytic extended to the larger set $\C \setminus \bigr(\sigma(H_\#) \cup \sigma(-H_\#)\bigr)$. In fact, it is shown in \cite{DDRF2023} that $\widehat{\chi_H}$ can be meromorphic extended to the domain 
\begin{align}
    \mathcal{D}_\Gamma \coloneqq \{z \in \C : |\mathrm{Re}(z)|< \Gamma \quad \mbox{or} \quad \mathrm{Im}(z) \neq 0 \}, \label{eq:domainGamma}
\end{align}
where $\Gamma \coloneqq \inf \sigma^{\mathrm{ess}}(H_\#)$ is the \emph{ionization threshold} of $H$. However, this extension is not maximal in general. For instance, some cancellations can occur due to the conjugation of the resolvent of $H_\#$ with the operator $B$. To precisely characterize the maximal meromorphic extension of $\widehat{\chi_H}$, let us define the \emph{single-particle excitation spectrum}\footnote{This terminology is inspired by the observation that, for one-body Hamiltonians $H = \sum_{j=1}^N 1 \otimes... h... \otimes 1$ with $h$ self-adjoint on $L^2(\Omega, d\mu)$ (e.g., for non-interacting systems), $\sigma_1(H_\#)$ corresponds to a subset of the excitation spectrum of the single-particle operator $h$.} of $H$ as
\begin{align}
    \sigma_1(H_\#) \coloneqq \{ \lambda \in \sigma(H) : B P^{H_\#}(B_\epsilon(\lambda)) \neq 0, \quad \mbox{for any $\epsilon >0$} \}, \label{eq:singlespectrumdef}
\end{align}
where $B_\epsilon(\lambda)\subset \C$ denotes the ball centered at $\lambda$ with radius $\epsilon$ and $P^{H_\#}(U)$ denotes the spectral projection of $H_\#$ on some Borel subset $U\subset \C$. We also define the discrete and essential parts of $\sigma_1(H_\#)$ as
\begin{align*}
    &\sigma^{\mathrm{disc}}_1(H_\#) \coloneqq \{ \lambda \in \sigma_1(H_\#) : \mbox{$\lambda$ isolated and $\rank BP^{H_\#}(\{\lambda\}) < \infty$ }\} \\
    &\sigma^\mathrm{ess}_1(H_\#) \coloneqq \sigma_1(H_\#) \setminus \sigma^{\mathrm{disc}}_1(H_\#).
\end{align*}
The first result of this paper is then the following.
\begin{theorem}[Maximal meromorphic extension] \label{thm:chiH} Let $H$ be a Hamiltonian satisfying Assumption~\ref{assump:main}. Then the maximal meromorphic (see Definition~\ref{def:meromorphic}) extension of the Fourier transform of $\chi_H$ is given by
\begin{align*}
\widehat{\chi_H} : &\mathcal{D} \rightarrow \mathcal{B}(L^2_{\rho_0},L^2_{\nicefrac{1}{\rho_0}}) \\
&z\mapsto \widehat{\chi_H}(z) = \sum_{\lambda \in \sigma_1^{\mathrm{disc}}(H_\#)} \frac{2\lambda}{z^2-\lambda^2}BP^{H_\#}(\{\lambda\}) B^\ast + B \int_{\sigma_1^{\mathrm{ess}}(H_\#)} \frac{2\lambda}{z^2-\lambda^2} \mathrm{d} P^{H_\#}_\lambda B^\ast .
\end{align*}
where
\begin{align}
    \mathcal{D} \coloneqq \{ z \in \C : \pm z \not \in \sigma_1^{\mathrm{ess}}(H_\#) \}.\label{eq:Ddef}
\end{align}
In particular, the set of poles of $\widehat{\chi_H}$ is given by $\sigma^{\mathrm{disc}}_1(H_\#)\cup \sigma_1^{\mathrm{disc}}(-H_\#)$. Moreover, these poles are all simple, and their rank is given by
\begin{align*}   \rank_\lambda(\widehat{\chi_H}) = \rank B P^{H_\#}(\{\lambda\}). 
\end{align*}
\end{theorem}

\begin{remark*}[Poles with infinite rank] In fact, any isolated point in $\sigma^{\mathrm{ess}}(H_\#)$ can also be seen as a pole of infinite rank of $\widehat{\chi_H}$. The reason for excluding such poles from the discrete spectrum is that compact perturbations of the operator $H_\#$ may turn these poles into an accumulation point of poles, thereby making these singularities no longer isolated. 
\end{remark*}

\begin{remark*}[Dark excitations] The complementary spectrum $\sigma(H_\#) \setminus \sigma_1(H_\#)$ corresponds to the dark excitations of $H$, i.e., the part of the spectrum that can not be obtained by shining light on the system and measuring the absorption cross-section.
\end{remark*}

\subsection{Well-posedness of the Dyson equation}

 Let us now turn to the solutions of the adiabatic Dyson equation \eqref{eq:Dysonadiabatic}. The first order of business is to agree on the underlying solution space. In LR-TDDFT, the goal of the Dyson equation is to approximate the DDRF of a system of interacting electrons via the DDRF of the equivalent non-interacting Kohn-Sham system. The equivalence is in the sense that the Hamiltonians of both the interacting and non-interacting systems have the same ground state density $\rho_0$. In particular, the DDRF of both systems should lie on the space of strongly continuous maps from $\R_+$ to $\mathcal{B}(L^2_{\rho_0}, L^2_{\nicefrac{1}{\rho_0}})$, denoted here by
\begin{align*}
    C_s\bigr(\R_+;\mathcal{B}(L^2_{\rho_0},L^2_{\nicefrac{1}{\rho_0}})\bigr).
\end{align*}
Therefore, it seems natural to study the well-posedness of the Dyson equation within this space. This choice is not unique, and we shall further motivate it later. Nevertheless, the Dyson equation is well-posed in this space under a compatible boundedness assumption on the adiabatic approximation of the Hxc-operator.
\begin{theorem}[Well-posedness of the Dyson equation]\label{thm:wellposedness} Let $F \in \mathcal{B}(L^2_{\nicefrac{1}{\rho_0}},L^2_{\rho_0})$ and $\chi_0 \in C_s\bigr(\R_+; \mathcal{B}(L^2_{\rho_0}, L^2_{\nicefrac{1}{\rho_0}})\bigr)$. Then, there exists a unique solution $\chi_F$ of the Dyson equation \eqref{eq:Dysonadiabatic}
in the space $C_s\bigr(\R_+; \mathcal{B}(L^2_{\rho_0}, L^2_{\nicefrac{1}{\rho_0}})\bigr)$. Moreover, the solution map
\begin{align*}
	\mathcal{S}_F : &C_s\bigr(\R_+; \mathcal{B}(L^2_{\rho_0}, L^2_{\nicefrac{1}{\rho_0}})\bigr) \rightarrow C_s\bigr(\R_+; \mathcal{B}(L^2_{\rho_0}, L^2_{\nicefrac{1}{\rho_0}})\bigr), \nonumber \\
	&\chi_0 \mapsto \chi_F
\end{align*}
is bijective.
\end{theorem}
The proof of Theorem~\ref{thm:wellposedness} is a standard application of Banach's fixed point theorem. For the details, we refer the reader to \cite[Section 3]{DDRF2023}, where the same theorem in a different function space is proved. Although the proof is rather simple, we show that Theorem~\ref{thm:wellposedness} guarantees the well-posedness of the Dyson equation for widely used adiabatic approximations of the Hxc-operator under general conditions on the ground state density $\rho_0$. In addition, the bijectivity of the solution map implies that, for any $F\in \mathcal{B}(L^2_{\nicefrac{1}{\rho_0}},L^2_{\rho_0})$, the DDRF of a Hamiltonian with ground state density $\rho_0$ can be obtained by solving the Dyson equation for a unique reference $\chi_0$. Of course, this does not guarantee that $\chi_0$ is the DDRF of a non-interacting Hamiltonian, a common premise of LR-TDDFT.

\subsection{The Dyson density-density response function}

Throughout this section, we assume that $\chi_F$ is the unique solution of the Dyson equation \eqref{eq:Dysonadiabatic} with $\chi_0 = \chi_H$ for some $H$ satisfying Assumption~\ref{assump:main}. Our goal is then to characterize these solutions and establish some of their fundamental properties. For this, the key ingredient is a representation formula for $\widehat{\chi_F}$ based on an operator version of the Casida matrix. More precisely, let us formally define the \emph{Casida} operator as
\begin{align*}
    \mathcal{C} \coloneqq H_\#^2 + 2H_\#^{\frac12} B^\ast F B H_\#^{\frac12}.
\end{align*} 
Then under the assumption that $F$ is symmetric (which is satisfied for physically relevant adiabatic approximations), we can apply the KLMN theorem (see Section~\ref{sec:preliminaries}) to properly define $\mathcal{C}$ as a semi-bounded self-adjoint operator on $\{\Psi_0\}^\perp$. Moreover, one can show (see Lemma~\ref{lem:CandC(z)}) that
\begin{align*}
    H_\#^{\frac12} (z^2- \mathcal{C})^{-1} H_{\#}^\frac12 , \quad \mbox{for $z^2 \not \in \sigma(\mathcal{C})$,}
\end{align*}
defines a bounded operator on $\{\Psi_0\}^\perp$. The key result of this paper is that the Fourier transform of $\chi_F$ is given by the conjugation of $B$ with the operator above. Precisely, we have
\begin{theorem}[Solution in the frequency domain] \label{thm:Casidarep} Let $F \in \mathcal{B}(L^2_{\nicefrac{1}{\rho_0}},L^2_{\rho_0})$ be symmetric, then the Fourier transform of $\widehat{\chi_F}$ is well-defined for $|\mathrm{Im}(z)| > \norm{B^\ast F B}$ and satisfies 
\begin{align}
    \widehat{\chi_F}(z) = 2B H_\#^{\frac12} (z^2-\mathcal{C})^{-1} H_\#^{\frac12}B^\ast. \label{eq:chiFFT}
\end{align}
\end{theorem}

We can now derive several properties of the solution $\chi_F$. For starters, we can take the inverse Fourier transform of eq.~\eqref{eq:chiFFT} to obtain the following representation formula for $\chi_F$ in the time domain. 
\begin{corollary}[Solution in the time domain]\label{cor:chiFtime}The solution $\chi_F$ is given by the formula
\begin{align}
    \chi_F(t) = -2\theta(t) t B H_\#^{\frac12} \mathrm{sinc}(t\sqrt{\mathcal{C}} ) H_\#^{\frac12} B^\ast \label{eq:timerepformula},
\end{align}
where $\mathrm{sinc}(s) = \sin(s)/s$ is the analytic sinc function.
\end{corollary}

\begin{remark*} As the sinc function has a power series with only quadratic terms, the function $\mathrm{sinc}(\sqrt{s})$ defines an entire function. Moreover, this function is bounded on any half-line $[\alpha,\infty)$. In particular, $\mathrm{sinc}(t\sqrt{\mathcal{C}} )$ uniquely defines a bounded operator for any self-adjoint operator bounded from below. The fact that $H_\#^{\frac12}\mathrm{sinc}(t\sqrt{\mathcal{C}}) H_\#^{\frac12}$ is also bounded in $\{\Psi_0\}^\perp$ will be shown in Section~\ref{sec:proofs}.
\end{remark*}


The above representation formula highlights some important features of the solution $\chi_F$. For instance, note that we can decompose 
\begin{align*}
    \sqrt{\mathcal{C}} = \sqrt{\mathcal{C}_+} + i \sqrt{\mathcal{C}_-}
\end{align*}
where $\mathcal{C}_+$ and $\mathcal{C}_-$ are the non-negative self-adjoint operators corresponding respectively to the positive real part (on $(0,\infty)$) and the non-negative part (on $(-\infty, 0]$) of the spectrum of $\mathcal{C}$. As these are mutually orthogonal operators, we have the decomposition
\begin{align*}
    \chi_F(t) = \underbrace{-2 \theta(t) t B H_\#^{\frac12} \mathrm{sinc}(t\sqrt{\mathcal{C}_+}) H_\#^{\frac12} B^\ast}_{=: \chi_F^+(t)} + \underbrace{-2 \theta(t) t B H_\#^{\frac12} \mathrm{sinc}(it\sqrt{\mathcal{C}_-}) H_\#^{\frac12} B^\ast}_{=: \chi_F^-(t)}.
\end{align*}
Consequently, if $0 \not \in \sigma(\mathcal{C})$, then the positive part $\chi_F^+$ is stable in the sense that it is uniformly bounded in time, as expected from a DDRF of an isolated quantum system. The negative part, on the other hand, grows exponentially fast with time. Nevertheless, note that, since the operator $\mathcal{C}$ is bounded from below, the exponential growth of $\chi_F^-$ is bounded by
\begin{align*}
    \norm{\chi_F^-(t)} \lesssim e^{t \sqrt{-\inf\sigma(\mathcal{C})}}.
\end{align*}
When $0 \in \sigma(\mathcal{C})$, the solution may contain a linearly growing part, corresponding to the spectral projection of $\mathcal{C}$ on $0$. In particular, if we gradually increase the strength of the adiabatic approximation by setting $F(\epsilon) = \epsilon F$, the point $\epsilon_0$ where the spectrum of the Casida operator reaches $0$ corresponds to a phase transition of the system.

The next corollary shows that the stability condition $\mathcal{C} > \delta$ can be re-stated in terms of the simpler operator
\begin{align}
    \mathcal{M} \coloneqq H_\# + 2 B^\ast F B. \quad \mbox{(with domain $D(H_\#)$).} \label{eq:Mdef}
\end{align}
\begin{corollary}[Stability condition]\label{cor:stability} Let $\mathcal{M}$ be the operator defined in \eqref{eq:Mdef}, then we have 
\begin{align*}
    0 \in \sigma(\mathcal{M}) \iff 0 \in \sigma(\mathcal{C})\quad\mbox{and}\quad \sigma(\mathcal{M}) \cap (-\infty,0) \neq \emptyset \iff \sigma(\mathcal{C})\cap (-\infty,0) \neq \emptyset.
\end{align*}
In particular, the solution $\chi_F$ is stable (in the sense described above) if and only if
\begin{align}
    \mathcal{M} \geq \delta \label{eq:positivestability}
\end{align}
for some $\delta >0$. Moreover, $\chi_F$ is a tempered distribution and $\widehat{\chi_F}$ admits an analytic extension to the upper half-plane if and only if
\begin{align}
    \mathcal{M} \geq 0. \label{eq:nonnegativestability}
\end{align}
\end{corollary}
Since $\chi_F$ is supposed to approximate the DDRF of another Hamiltonian (which is causal and bounded), the stability condition is expected to hold for physically relevant $F$. For the RPA, condition~\eqref{eq:positivestability} follows from the positivity of $F^{\rm RPA}$ (see eq.~\eqref{eq:RPAdef}) and the fact that $\inf \sigma(H_\#) >0$. For the ALDA, however, we are not aware of a general argument to prove that \eqref{eq:positivestability} or \eqref{eq:nonnegativestability} holds.
\begin{remark*}[Quantitative stability] A quantitative version of  Corollary~\ref{cor:stability} can be obtained via the min-max principle. Specifically, one can show that
\begin{align*}
    &\mathcal{M} \geq \delta >0\quad\mbox{implies}\quad \mathcal{C} \geq \delta \omega_1, \mbox{ and }\\
    &\mathcal{C} \geq \delta >0\quad \mbox{implies} \quad \mathcal{M} \geq \sqrt{\norm{B^\ast F B}^2 + \delta}- \norm{B^\ast F B},
\end{align*}
where $\omega_1 \coloneqq \inf \sigma(H_\#)>0$.
\end{remark*}
\begin{remark*}[Stability in the finite-dimensional case] A finite-dimensional analog of $\mathcal{M}$, and the associated stability condition, also appear in previous works where linear-response eigenvalue problems in finite dimensions are considered \cite{THOULESS1961,OLSEN1988,BaiCangLReigenvalueI2012,BaiCangLReigenvalueII2013,Brabec2015}. More precisely, $\mathcal{M}$ is an operator version of the matrix $M = A + B$ defined in the space of orbital pairs in \cite{Brabec2015}. 
\end{remark*}

Theorem~\ref{thm:Casidarep} also allows us to characterize the maximal meromorphic extension of $\widehat{\chi_F}$. This characterization requires a spectral gap assumption on $\mathcal{C}$ and resembles the characterization of $\widehat{\chi_H}$ given in Theorem~\ref{thm:chiH}. To state it precisely, let us define the single-particle spectrum of $\mathcal{C}$ as 
\begin{align*}
	\sigma_1(\mathcal{C}) \coloneqq \{ \lambda \in \C : B H_\#^{\frac12} P^{\mathcal{C}}(B_\epsilon(\lambda)) \neq 0 \quad \mbox{for any $\epsilon>0$ small} \},
\end{align*}
where $P^\mathcal{C}$ is the spectral projection of $\mathcal{C}$. As before, we define the discrete part of $\sigma_1(\mathcal{C})$ as the set of isolated points with $\rank BH_\#^{\nicefrac{1}{2}} P^{\mathcal{C}}(\{\lambda\}) < \infty$, and the essential part as the complement of the discrete part. Then, we have the following characterization.
\begin{corollary}[Maximal meromorphic extension of $\widehat{\chi_F}$]\label{cor:maximal} Suppose that $\sigma_1(\mathcal{C})$ has a spectral gap on the non-negative part of the spectrum, i.e., $[0,\infty) \not \subset \sigma_1(\mathcal{C})$. Then the maximal meromorphic extension of $\widehat{\chi_F}$ is given by
\begin{align*}
\widehat{\chi_F} : &\mathcal{D}_F \rightarrow \mathcal{B}(L^2_{\rho_0},L^2_{\nicefrac{1}{\rho_0}}) \\
&z\mapsto \widehat{\chi_F}(z) = \sum_{\lambda \in \sigma_1^{\mathrm{disc}}(\mathcal{C})} \frac{2}{z^2 - \lambda}BH_\#^{\frac12} P^{\mathcal{C}}(\{\lambda\}) H_\#^{\frac12} B^\ast + B H_\#^{\frac12}\int_{\sigma^{\mathrm{ess}}_1(\mathcal{C})} \frac{2}{z^2-\lambda} \mathrm{d} P^{\mathcal{C}}_\lambda H_\#^{\frac12} B^\ast,
\end{align*}
where
\begin{align}
    \mathcal{D}_F \coloneqq \{z \in \C : z^2 \not \in \sigma_1^{\mathrm{ess}}(\mathcal{C})\}.\label{eq:DFdef}
\end{align}
In particular, the set of poles of $\widehat{\chi_F}$ is given by $\{z\in \C: z^2 \in \sigma_1^{\mathrm{disc}}(\mathcal{C})\}$. 
\end{corollary}

\begin{remark*}[Poles rank and order] The non-zero poles of $\widehat{\chi_F}$ are all simple and satsify
\begin{align*}
\rank_\lambda (\widehat{\chi_F}) = \rank BH_\#^{\frac12} P^{\mathcal{C}}(\{\lambda\}).
\end{align*}
On the other hand if $0 \in \sigma_1^{\mathrm{disc}}(\mathcal{C})$, then $\widehat{\chi_F}$ has a pole of second order at the origin.
\end{remark*}
As a last consequence of Theorem~\ref{thm:Casidarep}, we show that under a compactness condition on $F$, the domain of maximal meromorphic continuation of $\widehat{\chi_F}$ agrees with the domain of maximal meromorphic continuation of $\widehat{\chi_H}$. As the proof of this result is more involved than the proof of the previous corollaries, we promote it to a theorem.
\begin{theorem}[Invariance of maximal domain]\label{thm:maximaldomain} Suppose that $F \in \mathcal{B}(L^2_{\nicefrac{1}{\rho_0}},L^2_{\rho_0})$ satisfies
\begin{align}
    B^\ast F B \in \mathcal{B}_\infty(D(H_\#),\{\Psi_0\}^\perp), \label{eq:compactnesscondition}
\end{align}
where $\mathcal{B}_\infty(D(H_\#),\{\Psi_0\}^\perp)$ denotes the set of compact operators from $D(H_\#)$ (endowed with the graph norm) to $\{\Psi_0\}^\perp$, then
\begin{align*}
	\mathcal{D}_F = \mathcal{D},
\end{align*}
where $\mathcal{D}$ and $\mathcal{D}_F$ are the maximal domains defined in  \eqref{eq:Ddef} and \eqref{eq:DFdef}, respectively.
\end{theorem}

The relevance of the above theorem is that for typical Schr\"odinger operators, the maximal domain of meromorphic continuation of $\widehat{\chi_H}$ is related to its ionization threshold via eq.~\eqref{eq:domainGamma}. Since the compactness condition~\eqref{eq:compactnesscondition} holds for standard adiabatic approximations (see Proposition~\ref{prop:compactnesscriterion} below), the above corollary implies that such approximations are not able to shift the ionization threshold of $H$ (which is the Kohn-Sham Hamiltonian in applications).

 \subsection{Applications}  \label{sec:applications}
 
 We now discuss some applications of the previous results in the context of LR-TDDFT. Throughout this section, we work within the quantum chemistry set-up where $\Omega = \R^3$ and the underlying single-particle space is the classical Lebesgue space $L^2(\R^3)$.

 In this setting, the typical Hamiltonians of interest (e.g., the molecular Hamiltonian) are Schr\"odinger operators 
 \begin{align*}
     H = -\Delta +V(r_1,...,r_N),
 \end{align*}
 where $\Delta$ is the Laplacian on $\R^{3N}$ and $V$ is some real-valued function that acts by multiplication. Under general assumptions on $V$ (e.g., $V$ is in the Kato class of $\R^{3N}$ \cite{SimonSchrSemigroup82}), the ground state density of $H$ is bounded whenever it exists. 
 In particular, the following criterion applies to many situations encountered in practice. (The proof is a straightforward application of H\"older's inequality.) 
 \begin{proposition}[Sufficient criterion for adiabatic approximations] \label{prop:adiabaticcriteria} Let $\rho_0 \in L^1(\R^3) \cap L^\infty(\R^3)$, and $F = F_1 + F_2$ satisfy
 \begin{align*}
     \norm{F_1 f}_{L^2(\R^3) +L^\infty(\R^3)} \lesssim \norm{f}_{L^1(\R^3) \cap L^2(\R^3)} \quad\mbox{and}\quad  |(F_2 f)(r)| \lesssim \rho_0(r)^{\delta} |f(r)|,
 \end{align*}
 for some $\delta\geq -1$. Then $F \in \mathcal{B}(L^2_{\nicefrac{1}{\rho_0}},L^2_{\rho_0})$.
 \end{proposition}
The above criterion is easily verified for the following adiabatic approximations:
 \begin{itemize}
 \item The random phase approximation (RPA). In the RPA, $F$ is given by
    \begin{align}
    \bigr(F^{\mathrm{RPA}}g\bigr)(r) = \int_{\R^3} \frac{g(r')}{|r-r'|} \mathrm{d}r' . \label{eq:RPAdef}
\end{align}
Thus from the Hardy-Littlewood-Sobolev (HLS) inequality, we conclude that $F^{\mathrm{RPA}} \in \mathcal{B}(L^2_{\nicefrac{1}{\rho_0}},L^2_{\rho_0})$. (In fact, we just need $\rho_0 \in L^\frac32(\R^3)$ here.) 
\item The Petersilka, Gossmann, and Gross approximation (PGG) \cite{PGG1996}. In the PGG approximation, the operator $F$ is given by
\begin{align*}
    \bigr(F^{\mathrm{PGG}}g\bigr)(r) = \bigr(F^{\mathrm{RPA}}g\bigr)(r) - \frac{1}{2} \int_{\R^3} \frac{|\gamma_H(r,r')|^2}{\rho_0(r)\rho_0(r')}\frac{g(r')}{|r-r'|}\mathrm{d}r',
\end{align*}
where $\gamma_H(r,r')$ is the ground state single-particle density matrix of the Hamiltonian associated to $\chi_H$. Hence, from the simple inequality $|\gamma_H(r,r')|^2 \leq \rho_0(r) \rho_0(r')$ and the HLS inequality, we also have $F^{\mathrm{PGG}} \in \mathcal{B}(L^2_{\nicefrac{1}{\rho_0}},L^2_{\rho_0})$ for any $\rho_0 \in L^{\frac32}(\R^3)$.
\item The adiabatic local density approximation (ALDA) \cite{ZangwillSoven80, marques2012fundamentals,ullrich2012time}. The ALDA is not a single approximation but rather a class of approximations. In the ALDA, the operator $F$ is given by
\begin{align*}
    \bigr(F^{\mathrm{ALDA}}_{\rho_0}g\bigr)(r)  = \bigr(F^{\mathrm{RPA}}g\bigr)(r) + \underbrace{\frac{\mathrm{d}^2 \bigr(\rho  \varepsilon^{\mathrm{HEG}}_{\mathrm{xc}}(\rho)\bigr)}{\mathrm{d}\rho^2} \biggr\rvert_{\rho = \rho_0(r)}}_{\coloneqq f_{\mathrm{xc}}^{\mathrm{HEG}}(\rho_0(r))} g(r),
\end{align*}
where $\varepsilon^{\mathrm{HEG}}_{\mathrm{xc}}(\rho) = \varepsilon^{\mathrm{HEG}}_{\mathrm{x}}(\rho) + \varepsilon^{\mathrm{HEG}}_{\mathrm{c}}(\rho)$ is the exchange-correlation energy per particle of the homogeneous electron gas. While the exchange part is known and given by
\begin{align}
    \varepsilon^{\mathrm{HEG}}_{\mathrm{x}}(\rho) = -C \rho^{\frac13}, \label{eq:exchangeHEG}
\end{align}
the correlation can only be approximated, which leads to different approximations of $F^{\mathrm{ALDA}}_{\rho_0}$. To see why such approximations also belong to $\mathcal{B}(L^2_{\nicefrac{1}{\rho_0}},L^2_{\rho_0})$, let us take the parametrization of $\varepsilon^{\mathrm{HEG}}_c$ introduced by Perdew and Wang \cite{PW92} as an example. The PW correlation approximation is
\begin{align}
    \varepsilon_{\mathrm{c}}^{\mathrm{PW92}}(\rho) = -2A(1+\alpha_1 \rho^{-\frac13}) \log \biggr(1+\frac{1}{\beta_1 \rho^{-\frac16} + \beta_2 \rho^{-\frac13} + \beta_3 \rho^{-\frac12}+ \beta_4 \rho^{-\frac{1+P}{3}} }\biggr), \label{eq:correlationPW92}
\end{align}
where $P = 1$ or $\frac34$, and $A, \alpha_1, \beta_1,\beta_2,\beta_3,\beta_4>0$ are parameters chosen to reproduce the asymptotics expansions of $\varepsilon_{\mathrm{c}}^{\mathrm{HEG}}$ in the low and high-density limits, and to fit data from quantum Monte Carlo simulations \cite{CeperleyQMCcorrelation} in the intermediate regime. Thus from \eqref{eq:exchangeHEG} and \eqref{eq:correlationPW92} (and some tedious calculations), one can check that
\begin{align*}
    |f_{\mathrm{xc}}^{\mathrm{HEG}}\bigr(\rho_0(r)\bigr)| \lesssim_{\norm{\rho_0}_{L^\infty}} \rho_0(r)^{\max\{\frac12, \frac{1+P}{3}\}-\frac43} \lesssim_{\norm{\rho_0}_{L^\infty}} \rho_0(r)^{-\frac56} .
\end{align*}
Therefore, $F^{\mathrm{ALDA}}_{\rho_0} \in \mathcal{B}(L^2_{\nicefrac{1}{\rho_0}},L^2_{\rho_0})$ for any bounded $\rho_0$. Other parametrizations of $\varepsilon^{\mathrm{HEG}}_c(\rho)$ also satisfy the above inequality as long as they reproduce (up to second derivatives) the asymptotic expansion of $\varepsilon_{\mathrm{c}}^{\mathrm{HEG}}$ in the low-density limit.
\end{itemize}
The above list contains the most common adiabatic approximations used in practice and is not exhaustive. Note also that all adiabatic approximations mentioned above are symmetric. In particular, Proposition~\ref{prop:adiabaticcriteria} guarantees that the solution formulas derived here apply to the Dyson equation with these approximations under the sole condition that the ground state density of the Kohn-Sham system is bounded.

\begin{remark*}[Absorption spectrum] It turns out that the ground state density of typical quantum systems is not only bounded but also decays exponentially fast at infinity \cite{agmon1975spectral,agmon2014lectures,SimonSchrSemigroup82}. In this case, the weighted density space $L^2_{\rho_0}$ contains functions that can grow exponentially fast at infinity. In particular, the polarizability tensor 
\begin{align*}
    A_{jk}(\omega) = \mathrm{Im}\inner{r_j, \widehat{\chi_H}(\omega) r_k}
\end{align*}
is a well-defined tempered distribution. Similarly, if the stability condition~\eqref{eq:nonnegativestability} holds, the polarizability tensor of the solution $\chi_F$ also defines a tempered distribution, which can be shown to display peaks (Dirac's delta) at the poles of $\widehat{\chi_F}$ (see \cite[Proposition 17 and 44]{GW0method2016}). In applications, these peaks correspond to approximations of the electronic excitation spectrum of many-body quantum systems. Quantifying how good these approximations are is an important open problem. 
\end{remark*}

As a final result, we present a simple sufficient criterion for the compactness property~\eqref{eq:compactnesscondition} that applies to the aforementioned adiabatic approximations. As a consequence, Theorem~\ref{thm:maximaldomain} shows that none of these adiabatic approximations are able to shift the ionization threshold of the Kohn-Sham system.
\begin{proposition}[Compactness criterion] \label{prop:compactnesscriterion} Suppose that $\rho_0 \in L^1(\R^3) \cap L^\infty(\R^3)$ and that the domain of $H$ is continuously embedded in the classical Sobolev space $\mathcal{H}^1(\Delta) = \{ \Psi \in L^2(\R^{3N}) : \int_{\R^{3N}} |\nabla \Psi|^2 \mathrm{d}r < \infty \}$. Then for any $F = F^{\mathrm{RPA}} + F_{\rho_0}$ with $F_{\rho_0}$ satisfying 
\begin{align}
    |F_{\rho_0} g (r)| \lesssim \rho_0(r)^{\delta} |g(r)| \quad \mbox{for all $g \in L^2_{\nicefrac{1}{\rho_0}}$ and some $\delta>-1$,} \label{eq:ALDAassump}
\end{align}
we have $B^\ast F B \in \mathcal{B}_\infty(D(H_\#), \{\Psi_0\}^\perp)$.
\end{proposition}

\begin{remark*}[Optimality of \eqref{eq:ALDAassump}] The condition $\delta>-1$ in Proposition~\ref{prop:compactnesscriterion} is optimal, as the following example shows. Let $N=1$ and define the adiabatic approximation $F_{\rho_0}$ as
\begin{align*}
\bigr(F_{\rho_0}f\bigr)(r) = c\rho_0(r)^{-1} f(r),
\end{align*}
for some $c \in \R$. Since $N=1$, the ground state density is simply $\rho_0(r) = |\Psi_0(r)|^2$ and the operators $B$ and $B^\ast$ reduce to
\begin{align*}
    B\Phi (r) = \overline{\Psi_0}(r) \Phi(r) - \inner{\Psi_0,\Phi} \rho_0(r) \quad\mbox{and}\quad (B^\ast f)(r) = f(r) \Psi_0(r) - \inner{\rho_0, f}\Psi_0(r).
\end{align*}
Thus we have $\mathcal{C} = H_\#^2 + 2c H_\#$, which implies that
\begin{align*}
    (\C\setminus \mathcal{D}_F)^2 = \sigma_1^{\mathrm{ess}}(\mathcal{C}) = \{\lambda^2 + 2 c\lambda : \lambda \in \sigma^{\mathrm{ess}}_1(H_\#) \} \neq \{\lambda^2 : \lambda \in \sigma_1^{\mathrm{ess}}(H_\#)\} = (\C\setminus \mathcal{D})^2,
\end{align*}
provided that $c \neq 0$ and $\sigma_1^{\mathrm{ess}}(H_\#)$ is non-empty. \end{remark*}

\textit{Outline of the paper.} We introduce some notation in the next paragraph. In the following section, we recall some well-known results about self-adjoint operators, their quadratic forms, and the associated Sobolev scale of spaces. These results are then used in Section~\ref{sec:proofs} to prove the main results of this paper. In the appendix, we briefly discuss the optimality of weighted density spaces as the domain and co-domain of the DDRF.

\subsection*{Notation} 

We denote the set of non-negative real numbers by $\R_+$. For $A$ and $B$ scalar quantities, $A \lesssim B$ means that there is an irrelevant positive constant $C$ such that $|A| \leq C |B|$. Occasionally, we also use $ A \lesssim_\epsilon B$ to indicate the dependence of the implicit constant on the additional parameter $\epsilon$.

Let $F$ be a Banach space, then we denote its norm by $\norm{\cdot}_F$, or simply by $\norm{\cdot}$ if the space is clear from the context. The set of linear continuous operators from $F$ to another Banach space $G$ is denoted by $\mathcal{B}(F,G)$; the set of compact operators is denoted by $\mathcal{B}_\infty(F,G)$. The operator norm is denoted by $\norm{T}_{F,G}$ or simply by $\norm{T}$ if the Banach spaces are clear from the context. The kernel and the range of $T$ are denoted, respectively, by $\ker T \subset F$ and $\ran T \subset G$. We also use $\rank T = \dim \ran T$ for the rank of $T$. The anti-dual of a Banach space $F$, i.e., the space of antilinear continuous functions from $F$ to $\C$ endowed with the operator norm is denoted by $F^\star$. For the Fourier transform of a function $f:\R \rightarrow F$, we use the physics convention 
     \begin{align} 
        \widehat{f}(\omega) = \int_\R f(t) e^{i t\omega} \mathrm{d} t .\label{eq:fourierconvention1d}
     \end{align}
For any Hilbert space $\mathcal{H}$, we adopt the convention that the inner-product $\inner{\cdot,\cdot}_{\mathcal{H}}$ is antilinear in the first variable and linear in the second.


For $1\leq p \leq \infty$, $L^p(\R^n)$ denotes the standard $L^p$ spaces with respect to the Lebesgue measure on $\R^n$. We also use $L^p(\R^n) + L^q(\R^n)$ and $L^p(\R^n) \cap L^q(\R^n)$ for the Banach spaces of Lebesgue-measurable functions with the norms
     \begin{align*} &\norm{f}_{L^p+L^q} \coloneqq \inf_{f = f_p+f_q} \{ \norm{f_p}_{L^p} + \norm{f_q}_{L^q} \} \quad \mbox{and}\quad \norm{f}_{L^p\cap L^q} \coloneqq \max\{ \norm{f}_{L^p},\norm{f}_{L^q}\}.
     \end{align*}
     
Lastly, we recall the definition of an operator-valued meromorphic function \cite[App. C]{dyatlov2019mathematical}.
\begin{definition}[Meromorphic operator-valued function] \label{def:meromorphic} Let $\mathcal{D} \subset \C$ be open, then we say that $K: \mathcal{D} \rightarrow \mathcal{B}(F,G)$ is a meromorphic operator-valued function if for any $z_0 \in \mathcal{D}$ there exists (i) a neighborhood $U_{z_0}\subset \mathcal{D}$ of $z_0$, (ii) finitely many finite rank operators $\{K_j\}_{j\leq M} \subset \mathcal{B}(F,G)$, and (iii) a holomorphic function $K_0 : U_{z_0} \rightarrow \mathcal{B}(F,G)$ such that
\begin{align*}
    K(z) = K_0(z) + \sum_{j=1}^M (z-z_0)^{-j} K_j \quad \mbox{for $z \in U_{z_0}$.}
\end{align*}
If $K_j \neq 0$ for some $j\geq 1$, we say that $z_0$ is a pole of $K$. In addition, if $K_j = 0$ for $j\geq 2$, then we say that $z_0$ is a simple pole of $K$ and define its rank as
\begin{align*}
    \rank_{z_0}(K) = \rank K_1.
\end{align*}
Moreover, we say that $\mathcal{D}$ is the maximal domain of $K$ if there exists no meromorphic extension of $K$ to a strictly larger connected domain. 
\end{definition}

\section{Mathematical background} \label{sec:preliminaries}
In this section we briefly recall some well-known facts about the scale of Sobolev spaces associated to self-adjoint operators and their quadratic forms. We also use this short recap to set-up some additional notation that will be used during the proofs from Section~\ref{sec:proofs}. The material presented here can be found in standard references such as \cite{ReedSimonI,ReedSimonII,ReedSimonIV}. 

\subsection{Sobolev scale of spaces}

Throughout this section, we let $A : D(A) \subset \mathcal{H} \rightarrow \mathcal{H}$ be some self-adjoint operator on a Hilbert space $\mathcal{H}$ satisfying the inequality
\begin{align*}
\inner{\Psi, A\psi} \geq  \norm{\Psi}^2\quad \mbox{for any $\Psi \in D(A)$.}
\end{align*}
(In the case $A \geq 1-c$ for some $c>0$, we replace $A$ by $A+c$ everywhere in the discussion below.)
Then by the spectral theorem, there exists a measure space $(X, \nu)$, an unitary map $U: \mathcal{H} \rightarrow L^2(X,d\nu)$, and a real-valued $\nu$-measurable function $a: X \rightarrow [1,\infty)$ such that 
\begin{align*}
    &U(D(A)) = L^2(X, a^2 d\nu) = \biggr\{ f \in L^2(X, d\nu) : \int_X |f(x)|^2 a(x)^2 d\nu(x) < \infty \biggr\},
\end{align*}
and $A$ acts on $L^2(X,d\nu)$ by multiplication by $a$, i.e., 
\begin{align*}
    &\bigr(UA U^\ast f\bigr)(x) = a(x)f(x).
\end{align*}
We then define the Sobolev spaces induced by $A$ as follows.
\begin{definition*}[Sobolev spaces] For $s\geq 0$, the Sobolev space of order $s$ is the set
\begin{align*}
    \mathcal{H}^s(A) \coloneqq \{\Psi \in \mathcal{H} : U\Psi \in L^2(X, a^s d\nu\bigr) \}
\end{align*}
endowed with the norm
\begin{align}
    \norm{\Psi}_{\mathcal{H}^s(A)}^2 \coloneqq \int_X |U\Psi(x)|^2 a(x)^s \mathrm{d} \nu(x) = \norm{ A^{\frac{s}{2}} \Psi}^2. \label{eq:Sobolevnorm}
\end{align}
Moreover, the negative Sobolev space of order $-s$ is defined as the anti-dual space of $\mathcal{H}^s(A)$, i.e., the set
\begin{align*}
    \mathcal{H}^{-s}(A) \coloneqq \mathcal{H}^s(A)^\star = \{ T: \mathcal{H}^s(A) \rightarrow \C \quad\mbox{antilinear and continuous} \}
\end{align*}
endowed with the operator norm. 
\end{definition*}
By the Riesz representation theorem, we can isometrically identify $\mathcal{H}$ with its anti-dual $\mathcal{H}^\star$ via the Riesz map
\begin{align*}
    \mathcal{R} : \mathcal{H} \rightarrow \mathcal{H}^\star \quad \Psi \mapsto \mathcal{R}\Psi = \inner{\cdot, \Psi}_{\mathcal{H}}
\end{align*}
In this way, we have a natural chain of dense inclusions
\begin{align}
    ... \subset \mathcal{H}^s(A) ... \subset \mathcal{H}^m(A)... \subset \mathcal{H} \overset{\mathcal{R}}{\cong} \mathcal{H}^\star ... \subset \mathcal{H}^{-m}(A) ... \subset \mathcal{H}^{-s}(A) ... \quad\mbox{( $s \geq m \geq 0$).}\label{eq:inclusionchain}
\end{align}
From \eqref{eq:Sobolevnorm}, the operator $A^s$ restricted to $\mathcal{H}^m(A)$ defines an isometric isomorphism between $\mathcal{H}^m(A)$ and $\mathcal{H}^{m-s}(A)$ for any , $0 \leq s \leq m$. Consequently, the adjoint map induces an isometric isomorphism from $\mathcal{H}^{s-m}(A)$ to $\mathcal{H}^{-m}(A)$. In particular, by the chain of inclusions in \eqref{eq:inclusionchain} and using the commutation relations
\begin{align*}
    A^s A^m = A^{s+m} = A^m A^s,
\end{align*}
the operator $A^s$ can be uniquely extended to a continuous\footnote{with respect to the inductive limit topology on $\mathcal{H}^{-\infty}(A)$.}  operator on the whole Sobolev scale of spaces
\begin{align*}
    A^s : \mathcal{H}^{-\infty}(A) \coloneqq \bigcup_{m \in \R} \mathcal{H}^m(A) \rightarrow \mathcal{H}^{-\infty}(A) \quad \mbox{(for any $s \in \R$).}
\end{align*}
Furthermore, the chain of inclusions in \eqref{eq:inclusionchain} also allow us to naturally define the operator 
\begin{align*}
    \mu - A : \mathcal{H}^s(A) \rightarrow \mathcal{H}^{s-2}(A)
\end{align*}
for any $s \in \R$ and $\mu \in \C$. A useful consequence of the spectral theorem is that the spectrum of $A$ is independent of the Sobolev scale used. More precisely, we have
\begin{lemma}[Inverse on Sobolev scale] \label{lem:Sobolevinverse} Let $\mu \in \C$, then the operator $\mu - A$ is invertible in $\mathcal{B}(\mathcal{H}^s(A),\mathcal{H}^{s-2}(A))$ for some $s\in \R$ if and only if it is invertible for every $s\in \R$. \end{lemma}
\begin{proof} First, we can use the Riesz representation theorem on $L^2(X,\mathrm{d}\nu)$ to extend the unitary map (given by the spectral theorem) $U: \mathcal{H} \rightarrow L^2(X, d\nu)$ to $U: \mathcal{H}^{-s}(A) \rightarrow L^2(X, a^{-s}d\nu)$ for any $s\geq 0$. More precisely, $UT \in L^2(X, a^{-s}\mathrm{d}\nu)$ is the unique function satisfying 
\begin{align*}
    T(U^\ast g) = \int_{X}\overline{g}(x) (UT)(x) \mathrm{d} \nu(x) \quad \mbox{for any $g \in L^2(X, a^s\mathrm{d}\nu)$.}
\end{align*}
Hence, the operator $U(\mu-A) U^{\ast}$ acts on $L(X,a^{-s}\mathrm{d}\nu)$ as pointwise multiplication by $\mu - a(x)$ for any $s\in \R$. One can now check that this operator is invertible if and only if $|\mu - a(x)| > \delta$ $\mu$-a.e. for some $\delta>0$, which is equivalent to $\mu-A$ being invertible on $\mathcal{H}$.
\end{proof}

Let us conclude this section with a distributional version of Stone's formula (cf. \cite[Theorem VII.13]{ReedSimonI}) that will be useful to establish the maximality of the meromorphic extensions from Theorem~\ref{thm:chiH} and Corollary~\ref{cor:maximal}.

\begin{lemma}[Stone's formula]\label{lem:Stonesformula} Let $A$ be a semi-bounded self-adjoint operator and $R_A(z) = (z - A)^{-1}$ denote the resolvent of $A$. Then for any $f \in C^\infty_c(\R)$ we have
\begin{align*}
    \lim_{\eta \ra 0^+} \int_{\R} f(\mu) \bigr(R_A(\mu - i\eta) - R_A(\mu+i\eta) \bigr) \mathrm{d} \mu = 2\pi i \int_{\R} f(\lambda) \mathrm{d} P^A_\lambda,
\end{align*}
where $P^A_\lambda$ is the spectral projection-valued measure of $A$ and the convergence is in the operator norm on $\mathcal{B}(\mathcal{H}^s(A),\mathcal{H}^{s+2}(A))$ for any $s \in \R$.
\end{lemma}

\begin{proof} Since $(\mu \pm i \eta -\lambda)^{-1}$ is uniformly bounded in $\lambda \in \R$ for $\eta >0$ fixed, from Fubini's theorem we have 
\begin{align*}
    \int_{\R} f(\mu) \bigr(R_A(\mu-i\eta) - R_A(\mu+i\eta)\bigr)\mathrm{d} \mu = 2i \int_{\R} \bigr(f \ast p_\eta)(\lambda) \mathrm{d} P^A_\lambda = 2i (f\ast p_\eta)(A)
\end{align*}
where $p_\eta(\mu) = \frac{\eta}{\mu^2 + \eta^2}$ is the Poisson kernel. As $g(A)$ commutes with $A$ for any continuous function $g$, it is enough to show that
\begin{align*}
	\lim_{\eta \ra 0^+} \norm{(A+c) (f\ast p_\eta (A) - \pi f(A))}_{\mathcal{B}(\mathcal{H})} = 0.
\end{align*}
This now follows from the continuity of the spectral calculus and the estimate
\begin{align*}
	&|\lambda + c| \bigr|(f\ast p_\eta)(\lambda) - \pi f(\lambda)| \lesssim \eta \int_{\R} |\omega|\bigr(|\widehat{f}(\omega)| + |\partial_\omega \widehat{f}(\omega)|\bigr) \mathrm{d} \omega \quad \mbox{(for any $\lambda \in \R$),}
\end{align*}
which can be shown by using the Fourier transform of the Poisson kernel $\widehat{p_\eta}(\omega) = \pi e^{-\eta |\omega|}$. \end{proof}

\subsection{Quadratic forms}

Let us now introduce the quadratic form associated to a semi-bounded operator $A$ and present the KLMN theorem \cite[Theorem X.17]{ReedSimonII}. For a proof, consult \cite{ReedSimonII}.
\begin{definition*}[Quadratic form] For a semi-bounded self-adjoint operator $A$ satisfying $A \geq 1-c$ for some $c \in \R$, the associated quadratic form is the sesquilinear map $q_A : \mathcal{H}^1(A) \times \mathcal{H}^1(A) \rightarrow \C$ defined as
\begin{align*}
    q_{A}(\Psi,\Phi) = \inner{(A+c)^{\frac12} \Psi, (A+c)^{\frac12} \Phi} - c \inner{\Psi,\Phi}.
\end{align*}
The Sobolev space of order $1$ is also called the form domain of $A$.
\end{definition*}
Note that, by definition,
\begin{align*}
    \Psi \mapsto q_A(\cdot, \Psi) = (A+c)^{\frac12} (A+c)^{\frac12}\Psi - c\Psi  = A \Psi \in \mathcal{H}^{-1}(A).
\end{align*}
Next, let $\beta : \mathcal{H}^1(A) \times \mathcal{H}^1(A) \rightarrow \C$ be another symmetric sesquilinear form. Then we say that $\beta$ is relatively bounded with respect to $q_A$ if there exists some $0< a < 1$  and $b >0$ such that
\begin{align}
    |\beta(\Phi,\Phi)| \leq a q_A(\Phi,\Phi) + b \norm{\Phi}^2 \quad \mbox{for any $\Phi \in \mathcal{H}^1(A)$.} \label{eq:smallquadraticform}
\end{align}
With this definition, we can now state the KLMN theorem, which is essentially a quadratic form version of the celebrated Kato-Rellich theorem. 
\begin{lemma}[KLMN theorem \cite{ReedSimonII}] \label{lem:KLMNthm} Let $A$ be a self-adjoint operator satisfying $A \geq 1-c$ and $\beta$ be a symmetric sesquilinear form on $\mathcal{H}^1(A)$ satisfying  \eqref{eq:smallquadraticform}. Then, there exists an unique self-adjoint operator $B$ with the same form domain as $A$ and satisfying
\begin{align}
    q_B = q_A + \beta .\label{eq:KLMNid}
\end{align}
Moreover, $B \geq (1-a)(1-c) -b$.
\end{lemma}

The proof of the KLMN theorem consists in using inequality~\eqref{eq:smallquadraticform} to show that the norms $q_B + \alpha \inner{\cdot, \cdot}$ and $\norm{ (A+c)^{\frac12} \cdot}$ are equivalent and exploit the one-to-one relation between semi-bounded closed quadratic forms and semi-bounded self-adjoint operatos. In particular, using interpolation theory one can show that
 \begin{align}
    \mathcal{H}^s(A) = \mathcal{H}^s(B) \quad \mbox{and} \quad \norm{(B+\alpha)^\frac{s}{2} \Phi} \sim \norm{(A+c)^{\frac{s}{2}} \Phi} \quad \mbox{ for any $\Phi \in \mathcal{H}^s(A)$,} \label{eq:equivalenceofSobolev}
\end{align}
provided that $-1\leq s \leq 1$. (The identity above means that these are the same subsets of $\mathcal{H}$ for $s\geq 0$, and not simply isometric.) 

\subsection{Reducing subspaces and block decomposition}

We now recall the definition of reducing subspaces for unbounded self-adjoint operators and the associated block decomposition. For the proofs of the results presented next, we refer to \cite[Section 1.4]{BookSchmudgen2012}. As before, we assume that $A: D(A) \subset \mathcal{H} \rightarrow \mathcal{H}$ is self-adjoint.

\begin{definition*}[Invariant and reducing subspaces] We say that some closed subspace $V \subset \mathcal{H}$ is an invariant subspace of $A$ if $A$ maps the intersection $D(A) \cap V$ to $V$. Moreover, we say that $V$ is a reducing subspace for $A$ if both $V$ and $V^\perp$ are invariant and the decomposition
\begin{align*}
	D(A) = \bigr(D(A) \cap V\bigr) \oplus \bigr(D(A) \cap V^\perp\bigr)
\end{align*}
holds.
\end{definition*}
The main motivation for introducing the notion of reducing spaces is the following well-known block decomposition.
\begin{theorem}[Block decomposition \cite{BookSchmudgen2012}]\label{thm:blockdecomposition} Suppose that $V$ is a reducing subspace for $A$, then 
\begin{align*}
	A \rvert_V : D(A) \cap V \rightarrow V \quad \mbox{and}\quad A \rvert_{V^\perp} : D(A) \cap V^\perp \rightarrow V^\perp
\end{align*}
are self-adjoint operators in $V$ and $V^\perp$, respectively, and we have $A = A\rvert_V \oplus A \rvert_{V^\perp}$. In particular, we have the spectral decomposition
\begin{align*}
\sigma(A) = \sigma(A\rvert_V) \cup \sigma(A\rvert_{V^\perp}),
\end{align*}
where $\sigma(A\rvert_V)$ and $\sigma(A\rvert_{V^\perp})$ are the spectra on $V$ and $V^\perp$, respectively.
\end{theorem}

While every invariant subspace is also reducing for bounded self-adjoint operators, this is no longer true for unbounded self-adjoint operators. For the latter, we shall use the following criterion for reducing subspaces.
\begin{lemma}[Criterion for reducing subspaces \cite{BookSchmudgen2012}] \label{lem:reducingcriterion} Let $V \subset \mathcal{H}$ be a closed subspace. Then $V$ is reducing for $A$ if and only if $V$ and $V^\perp$ are invariant for $A$ and the orthogonal projection on $V$ maps $D(A)$ to itself.
\end{lemma}

\section{Proofs} \label{sec:proofs}

In this section, we present the proof of the main results of this paper.
 
 \subsection{Proof of Theorem~\ref{thm:chiH}}
 
 The first step in the proof of Theorem~\ref{thm:chiH} is to show that the operators $B : \mathcal{H}_N \rightarrow L^2_{\nicefrac{1}{\rho_0}}$ and $B^\ast: L^2_{\rho_0}\rightarrow \mathcal{H}_N$ are bounded. For the boundedness of $B$, we can use the estimate
 \begin{align}
     B\Phi(r) &= N\int_{\Omega^{N-1}} \overline{\Psi_0(r,r_2,...,r_N)} \Phi(r,r_2,...,r_N)\mathrm{d}\mu(r_2)...\mathrm{d}\mu(r_N)  \leq \sqrt{\rho_0(r)}\sqrt{\rho_{\Phi}(r)} \label{eq:CSineq}
 \end{align}
 for any $\Phi \in \{\Psi_0\}^\perp$, and the fact that $B\Psi_0 = 0$. Moreover, by identifying $L^2_{\rho_0} \cong (L^2_{\nicefrac{1}{\rho_0}})^\star$ via the Riesz map on $L^2(\Omega, d\mu)$ and using the identity
 \begin{align*}
     \inner{B \Phi, g}_{L^2(\Omega,d\mu)} = \inner{\Phi, B^\ast g}_{\mathcal{H}_N},
 \end{align*}
 we conclude that $B^\ast \in \mathcal{B}(L^2_{\rho_0},\mathcal{H}_N)$ by duality. Hence, recalling that $\int \rho_{\Phi}(r) \mathrm{d}\mu(r) = N$ for any normalized $\Phi$, we have just proved the following proposition.
 \begin{proposition}[Boundedness on weighted density spaces]\label{prop:Bbound} The operators $B$ and $B^\ast$ satisfy
 \begin{align*}
    \norm{B}_{\mathcal{H}_N, L^2_{\nicefrac{1}{\rho_0}}} = \norm{B^\ast }_{L^2_{\rho_0},\mathcal{H}_N} \leq N.
\end{align*}
In particular, $\chi_H: \R \rightarrow \mathcal{B}(L^2_{\rho_0},L^2_{\nicefrac{1}{\rho_0}})$ is bounded and strongly continuous.
 \end{proposition}

We can now complete the proof of Theorem~\ref{thm:chiH}
\begin{proof}[Proof of Theorem~\ref{thm:chiH}]
Since $\chi_H$ is bounded and strongly continuous, its Fourier transform is well-defined as a tempered distribution. Moreover, since $\chi_H$ is causal (i.e., $\chi_H(t) = 0$ for $t\leq 0$), the Fourier transform is analytic on the upper half-plane $\{z \in \C: \mathrm{Im}(z) >0\}$. From the spectral Theorem and straightforward computation (see \cite[Section 2]{DDRF2023}), this analytic continuation is given by 
\begin{align*}
    \widehat{\chi_H}(z) = B \bigr((z-H_\#)^{-1} - (z+H_\#)^{-1}\bigr) B^\ast.
\end{align*}
Now note that, since the single-particle excitation spectrum $\sigma_1(H_\#)$ is closed (thus Borel measurable), we can decompose the resolvent of $H_\#$ in two terms:
\begin{align*}
    (z-H_\#)^{-1} = P^{H_\#}\bigr(\sigma_1(H_\#)\bigr) (z-H_\#)^{-1} + P^{H_\#}\bigr(\R\setminus \sigma_1(H_\#)\bigr)(z-H_\#)^{-1}.
\end{align*}
From the definition of $\sigma_1(H_\#)$, the second term vanishes when multiplied by $B$ on the left. In particular, the spectral theorem yields
\begin{align}
    \widehat{\chi_H}(z) = B \int_{\sigma_1(H_\#)} \frac{2 \lambda}{z^2-\lambda^2} \mathrm{d} P_\lambda^{H_\#} B^\ast .  \label{eq:spectralrep}
\end{align}
Next, observe that the spectral gap assumption on $H$ implies that the set $\{ z \in \C: \pm z \not \in \sigma_1(H_\#) \}$ is open and connected. Thus the right-hand side of \eqref{eq:spectralrep} defines the unique analytic extension of $\widehat{\chi_H}$ to this set. Moreover, for any isolated point $\lambda_0 \in \sigma_1(H_\#)$, we have
\begin{align*}
    \widehat{\chi_H}(z) = \frac{2\lambda_0}{z^2-\lambda_0^2} BP^{H_\#}(\{\lambda_0\}) B^\ast + B \int_{\sigma_1(H_\#) \setminus \{\lambda_0\}} \frac{2\lambda}{z^2-\lambda^2}\mathrm{d} P^{H_\#}_\lambda B^\ast,
\end{align*}
where the second term is analytic around $\lambda_0$. In particular, $\widehat{\chi_H}$ is meromorphic on
\begin{align}
    \mathcal{D} \coloneqq \{ z \in \C : \pm z \not\in \sigma_1^{\mathrm{ess}}(H_\#) \}.
\end{align}
From the identity $\rank S = \rank S S^\ast$, we also obtain the rank equality
\begin{align*}
    \rank_{\lambda_0}(\widehat{\chi_H}) = \rank B P^{H_\#}(\{\lambda_0\}) B^\ast = \rank B P^{H_\#}(\{\lambda_0\}) \quad \mbox{for any $\lambda_0 \in \sigma_1^{\mathrm{disc}}(H_\#)$.}
\end{align*}
To conclude the proof, we need to show that $\mathcal{D}$ is the maximal domain of meromorphic continuation of $\widehat{\chi_H}$. So suppose that $\widehat{\chi_H}$ is analytic around some $\lambda_0 \in \R$, then it is enough to show that $\lambda_0 \not \in \sigma_1(H_\#)$. Moreover, since $\widehat{\chi_H}(z) = \widehat{\chi_H}(-z)$, we can assume (without loss of generality) that $\lambda_0 \in \R_+$. Then, define
\begin{align*}
    T(z) \coloneqq \widehat{\chi_H}(z) - B (z+H_\#)^{-1} B^\ast.
\end{align*}
As the right-hand side is continuous close to $\lambda_0$, Stone's formula (see Lemma~\ref{lem:Stonesformula}) yields
\begin{align*}
    0 = \lim_{\eta \ra 0^+}\int_{\R} f(\lambda)\bigr(T(\lambda-i\eta) - T(\lambda + i\eta)\bigr) \mathrm{d} \lambda = 2 \pi i B\int_{\R} f(\lambda) \mathrm{d} P^{H_\#}_\lambda B^\ast
\end{align*}
for any $f\in C^\infty_c(\R)$ with support in $B_\epsilon(\lambda_0)$ for $\epsilon>0$ small enough. Choosing a sequence $f_n$ converging monotonically to the indicator function on $B_\epsilon(\lambda_0)\cap \R$ and using the strong convergence property of the spectral calculus  (see \cite[Theorem VIII.5.(d)]{ReedSimonI}), we find that $B P^{H_\#}(B_\epsilon(\lambda_0)) B^\ast = 0$. Therefore, $\lambda_0 \not \in \sigma_1(H_\#)$, which completes the proof. \end{proof}

\begin{remark*}[Regularity of $\widehat{\chi_H}$ along the continuous spectrum] If $H$ has compact resolvent (e.g., a Schr\"odinger operator with a trapping potential), then the maximal domain of meromorphic continuation is the whole complex plane. However, for typical Hamiltonians in electronic structure theory (e.g., the atomic and molecular Hamiltonians), the spectrum is divided into discrete and continuous parts \cite{ReedSimonIV}. 
In some exceptional cases, and for suitable $f$ and $g$, the regularity of the map $\omega \mapsto \inner{g,\widehat{\chi_H}(\omega) f}$ along the continuous spectrum can be rigorously studied (see \cite{dupuy:hal-03145143}) via the celebrated limiting absorption principle \cite{commutatormethodsbook,dyatlov2019mathematical}.
\end{remark*}

\subsection{Proof of Theorem~\ref{thm:Casidarep}}

We split the proof of Theorem~\ref{thm:Casidarep} into two parts. In the first part, we properly define the Casida operator and relate its resolvent to the inverse of the operator
\begin{align*}
    \mathcal{C}(z) = z^2 H_\#^{-1} - H_\# - 2 B^\ast F B.
\end{align*}
In the second part, we show that $\widehat{\chi_H}$ is given by the conjugation of $B$ with $\mathcal{C}(z)^{-1}$ via the convolution property of the Fourier transform and a well-known resolvent identity.

\subsubsection*{The Casida Operator}

To properly define the Casida operator, we consider the quadratic form $\beta : D(H_\#) \times D(H_\#) \rightarrow \C$ defined as
\begin{align*}
    &(\Psi, \Phi) \mapsto \beta(\Psi,\Phi) = \inner{H_{\#}^\frac12 \Psi, 2B^\ast F B H_\#^{\frac12} \Phi}.
\end{align*}
Recall that we assumed $F \in \mathcal{B}(L^2_{\nicefrac{1}{\rho_0}},L^2_{\rho_0})$ to be symmetric, and
\begin{align*}
    H_\# = H-\mathcal{E}_0\bigr\rvert_{\{\Psi_0\}^\perp} : D(H_\#) = D(H) \cap \{\Psi_0\}^\perp \rightarrow \{\Psi_0\}^\perp,
\end{align*}
where $\Psi_0$ and $\mathcal{E}_0$ are the ground state and ground state energy of $H$. Therefore, we can use the KLMN theorem to prove the following proposition.
\begin{proposition}[The Casida operator] \label{prop:casidadef} There exists an unique self-adjoint operator $\mathcal{C} : D(\mathcal{C}) \subset \{\Psi_0\}^\perp \rightarrow \{\Psi_0\}^\perp$ such that $\mathcal{H}^1(\mathcal{C}) = D(H_\#)$ and 
\begin{align*}
    q_{\mathcal{C}}(\Psi, \Phi) = \inner{H_\# \Psi, H_\# \Phi} + \beta(\Psi, \Phi)  \quad \mbox{for any $\Psi, \Phi \in D(H_\#)$.}
\end{align*}
Moreover, we have 
\begin{align}
    \mathcal{C} \geq \begin{dcases} \omega_1 (\omega_1 - 2\norm{B^\ast F B}) \quad &\mbox{if $\omega_1 \geq \norm{B^\ast F B}$,} \\
    -\norm{B^\ast F B}^2 \quad &\mbox{otherwise,} \end{dcases}\label{eq:infofC}
\end{align}
where $\omega_1 = \inf \sigma(H_\#) > 0$.
\end{proposition}
\begin{proof} Note that $q_{H_\#^2} = \inner{H\cdot, H\cdot}$ is the quadratic form of the positive operator 
\begin{align*}
    H_\#^2 : \mathcal{H}^4(H_\#) \subset \{\Psi_0\}^\perp \rightarrow \{\Psi_0\}^\perp.
\end{align*}
Consequently, to apply the KLMN theorem, we just need to check that $\beta$ is relatively bounded with respect to $q_{H_\#^2}$. For this, we can use Cauchy-Schwarz and Young's inequality to obtain
\begin{align}
    |\inner{H_\#^{\frac12} \Psi, 2B^\ast F B H_\#^{\frac12} \Psi}| &\leq 2\norm{H^{\frac12} \Psi}^2 \norm{B^\ast F B} \leq 2\norm{H_\# \Psi} \norm{\Psi} \norm{B^\ast F B} \label{eq:intermediateest} \\
    &\leq a \norm{H_\# \Psi}^2 + a^{-1} \norm{B^\ast F B}^2\norm{\Psi}^2  = a q_{H_\#^2}(\Psi,\Psi) + a^{-1} \norm{B^\ast F B}^2 \norm{\Psi}^2  \nonumber
\end{align}
for any $a>0$. To prove eq.~\eqref{eq:infofC} we can use eq.~\eqref{eq:intermediateest} and the simple estimate $\norm{H_\# \Phi} \geq \omega_1 \norm{\Phi}$.
\end{proof}

The next step is to relate the Casida operator to the operator
\begin{align}
    \mathcal{C}(z)  = z^2 H_\#^{-1} - H_\# -  2B^\ast F B \quad \mbox{with domain}\quad D\bigr(\mathcal{C}(z)\bigr) = D(H_\#). \label{eq:C(z)def}
\end{align}
Since $H_\#$ is self-adjoint and the operators $H_\#^{-1}$ and $B^\ast F B$ are both bounded and symmetric, the operator $\mathcal{C}(z)$ is closed and normal. Formally, the operator $\mathcal{C}$ is given by 
\begin{align*}
    H_\#^2 + 2H_\#^{\frac{1}{2}} B^\ast F B H_\#^{\frac{1}{2}}.
\end{align*}
We thus expect that 
\begin{align*}
 z^2 - \mathcal{C} = H_\#^{\frac12} \mathcal{C}(z) H_\#^{\frac12}
\end{align*}
in an appropriate sense. Rigorously clarifying this statement is the goal of the next lemma.

\begin{lemma}[The resolvent of $\mathcal{C}$]\label{lem:CandC(z)} Let $\mathcal{C}$ be the Casida operator defined according to Proposition~\ref{prop:casidadef} and $\mathcal{C}(z)$ be the operator defined above. Then $z^2 - \mathcal{C}$ is invertible if and only if $\mathcal{C}(z)$ is invertible. In this case, the operator $(z^2 - \mathcal{C})^{-1}$ maps $\mathcal{H}^{-1}(H_\#)$ onto $\mathcal{H}^3(H_\#)$ and we have
\begin{align}
    \mathcal{C}(z)^{-1} = H_\#^{\frac12} (z^2 - \mathcal{C})^{-1} H_\#^{\frac12}  \in \mathcal{B}\bigr(\{\Psi_0\}^\perp ,\mathcal{H}^2(H_\#)\bigr) \label{eq:inverseid}
\end{align}
\end{lemma}
\begin{proof} Let $\mathcal{H}^1(\mathcal{C})$ be the first Sobolev space associated with $\mathcal{C}$. Then by Lemma~\ref{lem:Sobolevinverse}, a point $z^2$ is in the resolvent set of $\mathcal{C}$ if and only if the extension $z^2 - \mathcal{C} : \mathcal{H}^1(\mathcal{C}) \rightarrow \mathcal{H}^{-1}(\mathcal{C})$ is invertible. Since $\mathcal{H}^1(\mathcal{C}) = D(H_\#) = \mathcal{H}^2(H_\#)$ (by Proposition~\ref{prop:casidadef}), this is equivalent to the map
 \begin{align*}
    z^2 - \mathcal{C} : &\mathcal{H}^2(H_\#) \rightarrow \mathcal{H}^{-2}(H_\#) \\
     &\Phi \mapsto z^2 \inner{\cdot, \Phi}  - q_{\mathcal{C}}(\cdot, \Phi) = z^2 \inner{\cdot, \Phi} - \inner{H_\# \cdot, H_\# \Phi} - \inner{H_\#^{\frac12} \cdot, B^\ast F B H_\#^{\frac12} \Phi}
 \end{align*}
 being bijective (by the closed graph theorem). However, the right-hand side of the above is equal to 
 \begin{align*}
    q_{\mathcal{C}(z)}(H_{\#}^{\frac12} \cdot, H_\#^{\frac12} \Phi)
 = H_\#^{\frac 12} \mathcal{C}(z) H_\#^{\frac12} \Phi,
 \end{align*}
 where $\mathcal{C}(z)$ is the unique extension of $\mathcal{C}(z)$ in $\mathcal{B}(\mathcal{H}^1(H_\#),\mathcal{H}^{-1}(H_\#))$. In particular, we have
 \begin{align}
    z^2 - \mathcal{C} = H_\#^{\frac12} \mathcal{C}(z) H_\#^{\frac12} \quad\mbox{as a map from $\mathcal{H}^2(H_\#)$ to $\mathcal{H}^{-2}(H_\#)$.} \label{eq:asmap}
\end{align}
Thus since $H_\#^{\frac12} : \mathcal{H}^s(H_\#) \rightarrow \mathcal{H}^{s-1}(H_\#)$ is an isomorphism for every $s\in \R$, we conclude that $z^2 - \mathcal{C}$ is invertible on $\mathcal{B}(\mathcal{H}^2(H_\#),\mathcal{H}^{-2}(H_\#))$ if and only if $\mathcal{C}(z)$ is invertible on $\mathcal{B}(\mathcal{H}^1(H_\#), \mathcal{H}^{-1}(H_\#))$. 

To show \eqref{eq:inverseid}, we note that if either $\mathcal{C}(z)$ or $z^2-\mathcal{C}$ is invertible, then from \eqref{eq:asmap} we obtain
\begin{align}
    (z^2 - \mathcal{C})^{-1} = H_{\#}^{-\frac12} \mathcal{C}(z)^{-1} H_{\#}^{-\frac12} \quad \mbox{in $\mathcal{B}\bigr(\mathcal{H}^{-2}(H_\#),\mathcal{H}^{2}(H_\#)\bigr)$.} \label{eq:almostinverse}
\end{align}
Since $H_\#^{-\frac12} \mathcal{C}(z)^{-1}$ maps $\{\Psi_0\}^\perp$ bijectively to $\mathcal{H}^3(H_\#)$, eq.~\eqref{eq:almostinverse} implies that
\begin{align*}
    (z^2 - \mathcal{C}) \Phi \in \mathcal{H}^3(H_\#) \quad \mbox{for any $\Phi \in \mathcal{H}^{-1}(H_\#)$.}
\end{align*}
Eq.~\eqref{eq:inverseid} now follows by multiplying  eq.~\eqref{eq:almostinverse} by $H_\#^{\frac12}$ on the left and on the right. \end{proof}

\subsubsection*{Proof of Theorem~\ref{thm:Casidarep}}

We are now in position to prove Theorem~\ref{thm:Casidarep}. For this, we shall use the following well-known resolvent identity.
\begin{lemma}[First resolvent identity]\label{lem:resolventid} Let $\mathcal{C}(z)$ be the operator defined in \eqref{eq:C(z)def}. Then, if the operators $\mathcal{C}(z)$ and $(z^2 H_\#^{-1} - H_\#) : D(H_\#) \rightarrow \{\Psi_0\}^\perp$ are both invertible, we have 
\begin{align}
    \mathcal{C}(z)^{-1} - (z^2 H_\#^{-1} - H_\#)^{-1} &=  2 \mathcal{C}(z)^{-1} B^\ast F B (z^2 H_\#^{-1} - H_\#)^{-1}\label{eq:firstresolventid1} \\
    &=  2 (z^2 H_\#^{-1} - H_\#)^{-1} B^\ast F B \mathcal{C}(z)^{-1}. \label{eq:firstresolventid2}
\end{align}
\end{lemma}

\begin{proof}[Proof of Theorem~\ref{thm:Casidarep}]First, note that from the Dyson equation~\eqref{eq:Dysonadiabatic}, the simple estimate $\sup_{t \in \R} \norm{\chi_H(t)} \leq \norm{B}\norm{B^\ast} = \norm{B}^2$, and Gronwall's inequality we have
\begin{align*}
    \norm{\chi_F(t)} \leq \norm{B}^2 e^{\norm{B}^2 \norm{F} t} \quad\mbox{for $t\geq 0$.}
\end{align*}
In particular, the Fourier transform is well-defined and analytic for $\mathrm{Im}(z) > \norm{B}^2 \norm{F}$. In this case, by the convolution property of the Fourier transform, we have
\begin{align}
    \widehat{\chi_F}(z) = \widehat{\chi_H}(z) + \widehat{\chi_H}(z) F \widehat{\chi_F}(z),  \label{eq:frequencyDyson}
\end{align}
which is the frequency version of the Dyson equation. 

The idea now is to find a representation formula for the inverse of the dielectric operator 
\begin{align*}
    \varepsilon(z) \coloneqq 1-\widehat{\chi_H}(z)F = 1- 2B(z^2H_\#^{-1} - H_\#)^{-1} B^\ast F,
\end{align*}
where the second equality comes from the expression 
\begin{align*}
    \widehat{\chi_H}(z) = B H_\#^{\frac12}\bigr((z-H_\#)^{-1} - (z+H_\#)^{-1}\bigr) B^\ast = 2B (z^2H_\#^{-1} - H_\#)^{-1} B^\ast.
\end{align*}
For this, we can now use the resolvent identities from Lemma~\ref{lem:resolventid}. Precisely, let $\mathcal{C}(z)$ be defined as in eq.~\eqref{eq:C(z)def}.  Thus since $\mathcal{C}$ is self-adjoint and bounded from below by $-\norm{B^\ast F B}^2$,  Lemma~\ref{lem:CandC(z)} guarantees that $\mathcal{C}(z)$ is invertible for any $z$ with $|\mathrm{Im}(z)| \geq \norm{B^\ast F B}$. Similarly, the operator \[z^2 H_\#^{-1} - H_\# = H_\#^{-\frac12} (z^2 - H_\#^2) H_\#^{-\frac12}\]
is invertible for any $z$ with $z^2 \not \in (0,\infty)$. We now claim that 
\begin{align*}
    \varepsilon(z)^{-1} = 1 + 2 B\mathcal{C}(z)^{-1} B^\ast F \quad \mbox{for any $z$ with $\mathrm{Im}(z)$ large.}
\end{align*}
Indeed, from \eqref{eq:firstresolventid1} we have
\begin{multline}
    \bigr(1+2 B \mathcal{C}(z)^{-1} B^\ast F\bigr) \varepsilon(z)\\
    = 1+ 2B\underbrace{\biggr(\mathcal{C}(z)^{-1} - (z^2 H_\#^{-1} - H_\#)^{-1} - 2\mathcal{C}(z)^{-1} B^\ast F B (z^2 H_\#^{-1} - H_\#)^{-1}\biggr)}_{=0} B^\ast F = 1.
\end{multline}
On the other hand, eq.~\eqref{eq:firstresolventid2} implies that $\varepsilon(z)(1+ 2 B \mathcal{C}(z)^{-1} B^\ast) = 1$ as well, which proves our claim. To conclude, we note that from the (frequency) Dyson equation~\eqref{eq:frequencyDyson} we have
\begin{align*}
    \widehat{\chi_F}(z) &= \varepsilon(z)^{-1} \widehat{\chi_H}(z) = \bigr(1 + 2B \mathcal{C}(z)^{-1} B^\ast F\bigr) 2 B (z^2 H_\#^{-1} - H_\#)^{-1} B^\ast \\
    &= 2B\biggr( (z^2H_\#^{-1} - H_\#)^{-1} + 2\mathcal{C}(z)^{-1} B^\ast F B (z^2 H_\#^{-1} + H_\#)^{-1}\biggr)B^\ast \overset{\eqref{eq:firstresolventid1}}{=} 2B \mathcal{C}(z)^{-1} B^\ast.
\end{align*}
Thus Theorem~\ref{thm:Casidarep} follows from the equation above and Lemma~\ref{lem:CandC(z)}. \end{proof}

\subsection{Proof of Corollaries~\ref{cor:chiFtime} to \ref{cor:maximal}}

We now present the proofs of the corollaries from Theorem~\ref{thm:Casidarep}. We start with the proof of Corollary~\ref{cor:chiFtime}.

\begin{proof}[Proof of Corollary~\ref{cor:chiFtime}]
First, note that the power series of the sinc function is composed only of even terms. Thus the function 
\begin{align*}
    \lambda \in \C \mapsto f(\lambda) \coloneqq \mathrm{sinc}(\sqrt{\lambda})
\end{align*}
defines an analytic function in the whole complex plane. Moreover, this function is bounded on any half-line $\{\lambda \in \R : \lambda \geq -c \}$ for $c>0$. In particular, the operator $f( t^2 \mathcal{C})$ defined via the spectral theorem belongs to $\mathcal{B}(\{\Psi_0\}^\perp)$ for any $t \in \R$. In fact, by taking any function $g: \R \rightarrow [1,\infty)$ satisfying $g(\lambda) \sim \sqrt{\lambda}$ for $\lambda$ big we can re-write
\begin{align}
f(t^2\mathcal{C}) = g(\mathcal{C})^{-\frac12} h(t,\mathcal{C}) g(\mathcal{C})^{-\frac12} ,\label{eq:breakup}
\end{align}
where $h(t,\lambda) = g(\lambda) f(t^2\lambda)$ is bounded on $\sigma(\mathcal{C})$. So by recalling that $\mathcal{H}^s(\mathcal{C}) = \mathcal{H}^{2s}(H_\#)$ for any $-1\leq s \leq 1$ (because $\mathcal{H}^1(\mathcal{C}) = D(H_\#)$ with equivalence of norms), eq.~\eqref{eq:breakup} implies that
\begin{align*}
    f(t^2 \mathcal{C}) \in \mathcal{B}(\mathcal{H}^{-1}(H_\#),\mathcal{H}^1(H_\#)).
\end{align*}
Therefore, the operator-valued map
\begin{align}
	t  \in \R \mapsto \widetilde{\chi}(t) \coloneqq -2\theta(t) t B^\ast H_\#^{\frac12} f(t^2 \mathcal{C}) H_\#^{\frac12} B 
\end{align}
defines a strongly continuous family of operators in $\mathcal{B}(L^2_{\rho_0},L^2_{\nicefrac{1}{\rho_0}})$. To conclude the proof, we can use the identity
\begin{align*}
	\int_0^\infty \frac{\sin(\lambda t)}{\lambda} e^{i(\omega+i\eta)t} \mathrm{d} t = \frac{-1}{(\omega+i\eta)^2 - \lambda^2} \quad \mbox{(valid for any $\eta > |\mathrm{Im}(\lambda)|$)}
\end{align*}
to show that the Fourier transform of $\chi_F(t) e^{-\eta t}$ and $\widetilde{\chi}(t) e^{-i\eta t}$ coincide for $\eta$ large enough, and therefore, $\widetilde{\chi}(t) = \chi_F(t)$ for every $t \in \R$. (Note that equality holds everywhere because both functions are strongly continuous.)
\end{proof} 

Next, let us turn to the proof of the stability criterion.

\begin{proof}[Proof of Corollary~\ref{cor:stability}] First, note that
\begin{align*}
    \mathcal{M} = H_\# + 2 B^\ast F B = -\mathcal{C}(0),
\end{align*}
where $\mathcal{C}(z)$ is the operator defined in \eqref{eq:C(z)def}. In particular, by Lemma~\ref{lem:CandC(z)} we have $0 \in \sigma(\mathcal{C})$ if and only if $0 \in \sigma(\mathcal{M})$. Moreover, from the proof of Lemma~\ref{lem:CandC(z)} we know that
\begin{align*}
    q_{\mathcal{C}}(\Psi,\Psi) = q_{\mathcal{M}}(H_{\#}^{\frac12} \Psi, H_\#^{\frac12} \Psi)\quad \mbox{for any $\Psi \in D(H_\#)$.}
\end{align*}
Since a self-adjoint operator is non-negative if and only if its quadratic form is non-negative, the above identity (and the fact that $H_\#^{\frac12}: D(H_\#) \rightarrow \mathcal{H}^1(H_\#)$ is bijective) implies that $\mathcal{C}\geq 0$ if and only if $\mathcal{M} \geq 0$, which completes the proof.\end{proof}

Lastly, we present the proof of Corollary~\ref{cor:maximal}.
\begin{proof}[Proof of Corollary~\ref{cor:maximal}] Let us define
\begin{align*}
    \widetilde{\chi}(z) \coloneqq B H_\#^{\frac12} \int_{\sigma_1(\mathcal{C})} \frac{2}{z^2-\lambda} \mathrm{d} P^{\mathcal{C}}_\lambda H_\#^{\frac12} B^\ast.
\end{align*}
Then by multiplying and dividing by $g(\mathcal{C})^{\frac12}$ as we did in eq.~\eqref{eq:breakup}, we see that $\widetilde{\chi}(z)$ is bounded on $\mathcal{B}(L^2_{\rho_0},L^2_{\nicefrac{1}{\rho_0}})$ for any $z^2 \not \in \sigma_1(\mathcal{C})$. Moreover, $\widetilde{\chi}(z)$ is meromorphic on 
\begin{align}
    \mathcal{D}_F = \{ z \in \C : z^2 \not \in \sigma_1^{\mathrm{ess}}(\mathcal{C})\},
\end{align}
and its poles are located at $\{z \in \C : z^2 \in \sigma_1^{\mathrm{disc}}(\mathcal{C})\}$. From the spectral gap assumption on $\sigma_1(\mathcal{C})$, the open set $\mathcal{D}_F$ is connected. Thus since $\widehat{\chi_F}(z) = \widetilde{\chi}(z)$ for $\mathrm{Im}(z)$ big enough, the function $\widetilde{\chi}$ is the unique meromorphic extension of $\widehat{\chi_F}$ to $\mathcal{D}_F$. 

To show that this extension is maximal, we can now use Stone's formula as we did in the proof of Theorem~\ref{thm:chiH}.  Precisely, let $\mu_0 \in \C$ with $\mu_0^2\in \R$ and suppose that $\widehat{\chi_F}$ can be analytic extended to a neighborhood $U$ of $\mu_0$. Then it suffices to show that $\mu_0^2 \not \in \sigma_1(\mathcal{C})$. In the case $\mu_0^2 >0$, we can assume that $\overline{U}$ does not intersect the imaginary axis and define
\begin{align}
    \alpha^{\pm}(\mu,\eta) = - \mu + \sqrt{\mu^2 \pm i \eta}, \label{eq:alphadef}
\end{align}
where we choose the branch of the square root such that $\lim_{\eta \ra 0}\alpha^{\pm}(\eta,\mu) = 0$. Note that $\alpha^{\pm}$ is continuous in a neighborhood of $\eta =0$ and $\mu \in U$. Thus from the continuity of $\widehat{\chi_F}$ on $U$ and Stone's formula in Lemma~\ref{lem:Stonesformula} we have
\begin{align*}
    0&= \lim_{\eta \ra 0^+} \int_{\R_+} 2\mu f(\mu^2) \biggr(\widehat{\chi_F}\bigr(\mu+\alpha^+(\mu,\eta)\bigr) - \widehat{\chi_F}\bigr(\mu+\alpha^-(\mu,\eta)\bigr)\biggr) \mathrm{d} \mu \\
    &= \lim_{\eta \ra 0^+} BH_\#^{\frac12} \int_{\R_+} 2\mu f(\mu^2) \bigr((\mu^2 + i\eta - \mathcal{C})^{-1} - (\mu^2-i\eta - \mathcal{C})^{-1}\bigr) \mathrm{d} \mu  H_\#^{\frac12} B^\ast \\
    &= BH_\#^{\frac12} \int_\R f(\mu) \mathrm{d}\mathrm{P}^{\mathcal{C}}_\mu H_\#^{\frac12} B^\ast
\end{align*}
for any $f \in C^\infty_c(\R)$ with support on $\{ \lambda^2 \in U\cap \R \}$. As in the proof of Theorem~\ref{thm:chiH}, we can now use the strong-convergence property of the functional spectral calculus to conclude that $\mu_0^2 \not \in \sigma_1(\mathcal{C})$. 

In the case $\mu_0^2<0$, we can choose the neighborhood $U$ so that $\overline{U}$ does not intersect the real axis. By defining $\alpha^\pm(\mu,\eta)$ via \eqref{eq:alphadef} with the opposite branch of the square root, and using similar arguments, one can prove that $\mu_0^2 \not \in \sigma_1(\mathcal{C})$. Finally, if $\mu_0 =0$, then $\mu_0$ is at most an isolated point in $\sigma_1(\mathcal{C})$ by the preceding arguments. In this case, however, $\mu_0$ would be a pole of $\widehat{\chi_F}$, which is not possible because we assumed that $\widehat{\chi_F}$ is bounded around $\mu_0$. Therefore, $\mu_0^2 \not \in \sigma_1(\mathcal{C})$, which concludes the proof. \end{proof}

\subsection{Proof of Theorem~\ref{thm:maximaldomain}}

For the proof of Theorem~\ref{thm:maximaldomain}, it is more convenient to work with the single-particle excitation spectrum of $H_\#^2$ than of  $H_\#$, where the former is defined as 
\begin{align*}
\sigma_1(H_\#^2) \coloneqq \{\lambda \in \R: BP^{H_\#^2}(B_\epsilon(\lambda)) \neq 0 \quad \mbox{for any $\epsilon>0$ small} \}.
\end{align*}
More precisely, note that from the definitions of $\mathcal{D}$ and $\mathcal{D}_F$ (see \eqref{eq:Ddef} and \eqref{eq:DFdef}), the proof of Theorem~\ref{thm:maximaldomain} is done if we show that
\begin{align}
    \sigma_1^{\mathrm{ess}}(\mathcal{C}) = \sigma_1^{\mathrm{ess}}(H_\#^2),\label{eq:essentialspecequivalence}
\end{align}
where the essential part of $\sigma_1(H_\#^2)$ is defined in the same way as for $H_\#$. For this, the first step is the following lemma. 

\begin{lemma}[Reducing subspaces of $H_\#^2$ and $\mathcal{C}$]\label{lem:reducingC} Let 
\begin{align}
    V_H \coloneqq P^{H_\#^2}(\sigma_1(H_\#^2)) \quad \mbox{and}\quad V_{\mathcal{C}} \coloneqq P^{\mathcal{C}}(\sigma_1(\mathcal{C})). \label{eq:reducingsubspaces}
\end{align}
Then $V_H$ and $V_{\mathcal{C}}$ are reducing subspaces for both $H_\#^2$ and $\mathcal{C}$ and we have 
\begin{align}
    \sigma_1(\mathcal{C}) \subset \sigma(\mathcal{C}\rvert_{V_H})\quad\mbox{and}\quad \sigma_1(H_\#^2) \subset \sigma(H_\#^2\rvert_{V_\mathcal{C}}). \label{eq:inclusions}
\end{align}
\end{lemma}
\begin{proof} From the spectral theorem, $V_H$ and $V_{\mathcal{C}}$ are clearly reducing subspaces for $H_\#^2$ and $\mathcal{C}$, respectively. Let us then show that $V_H$ is reducing for $\mathcal{C}$. First, we claim that 
\begin{align}
    D(H_\#^2) \cap V_H^\perp = D(\mathcal{C}) \cap V_H^\perp \quad \mbox{and} \quad H_\#^2\rvert_{D(H_\#^2)\cap V_H^\perp} = \mathcal{C}\rvert_{D(H_\#^2)\cap V_H^\perp}  .\label{eq:domainid}
\end{align} 
To prove this claim, first note that there exists a sequence $\{(\lambda_j,\epsilon_j)\}_{j\in\N} \subset \R\times (0,1]$
such that
\begin{align*}
    BP^{H_\#^2}(B_{\epsilon_j}(\lambda_j)) = 0\quad \mbox{for any $j$ and } \lim_{m\ra \infty} \sum_{j=1}^m P^{H_\#^2}(B_{\epsilon_j}(\lambda_j)) = P^{H_\#^2}(\R\setminus \sigma_1(H_\#^2)) = : P_{V_H^\perp}
\end{align*}
in the strong sense. As a consequence,
\begin{align}
    B P_{V_H^\perp} = 0 \quad \mbox{and}\quad  P_{V_H^\perp} B^\ast = 0\label{eq:BVH}
\end{align}
Next, recall that
\begin{align*}
q_{\mathcal{C}}(\cdot,\cdot) = q_{H_\#^2}(\cdot,\cdot) + 2\inner{H_\#^{\frac12}\cdot, B^\ast F B H_\#^{\frac12} \cdot}. 
\end{align*}
Thus since $P_{V_H^\perp} = P^{H_\#^2}(\R\setminus \sigma_1(H_\#^2))$ commutes with $H_\#^{\frac12}$, eq.~\eqref{eq:BVH} implies that
\begin{align}
    q_{\mathcal{C}}(\Psi, P_{V_H^\perp} \Phi) = q_{H_\#^2}(\Psi, P_{V_H^\perp} \Phi) = q_{H_\#^2}(P_{V_H^\perp} \Psi, \Phi) =q_{\mathcal{C}}(P_{V_H^\perp}\Psi, \Phi)\label{eq:quadraticcancel}
\end{align}
for any $\Psi,\Phi \in D(H_\#)$. From the first identity in \eqref{eq:quadraticcancel}, we see that \eqref{eq:domainid} holds. From the last identity in \eqref{eq:quadraticcancel}, we find that $P_{V_H^\perp}$ maps $D(\mathcal{C})$ to itself and $V_H$ and $V_H^\perp$ are invariant subspaces for $\mathcal{C}$. We thus conclude from Lemma~\ref{lem:reducingcriterion} that $V_H$ is a reducing subspace for $\mathcal{C}$.

To prove the first identity in \eqref{eq:reducedessentialspec}, we now let $\lambda \not \in \sigma(\mathcal{C}\rvert_{V_H})$ and show that $\lambda \not \in \sigma_1(\mathcal{C})$. So first, as $\sigma(\mathcal{C}\rvert_{V_H})$ is closed, we can find $\epsilon>0$ such that $B_{\epsilon}(\lambda) \cap \sigma(\mathcal{C}\rvert_{V_H})= \emptyset$. Thus from the block decomposition in Theorem~\ref{thm:blockdecomposition}, we must have $\ran P^{\mathcal{C}}(B_\epsilon(\lambda))
 \subset V_H^\perp$. But since $P_{V_H^\perp}$ commutes with $H_\#$, we find that
\begin{align*}
    B H_\#^{\frac12} P^{\mathcal{C}}(B_\epsilon(\lambda)) = B H_\#^{\frac12} P_{V_H^\perp} P^\mathcal{C}(B_\epsilon(\lambda)) = B P_{V_H^\perp} H_\#^{\frac12} P^\mathcal{C}(B_\epsilon(\lambda)) \overset{\eqref{eq:BVH}}{=} 0,
\end{align*}
which implies that $\lambda \not \in \sigma_1(\mathcal{C})$ by definition. 

Finally, to prove that $V_\mathcal{C}$ is reducing for $H_\#^2$ and that the second inclusion in \eqref{eq:inclusions} holds, we can reverse the roles of $\mathcal{C}$ and $H_\#^2$. More precisely, we let $\alpha>0$ be big enough (e.g., $\alpha > \norm{B^\ast F B}^2 +1$ will do) and note that
\begin{align*}
    &q_{H_\#^2}(\cdot,\cdot) = q_{\mathcal{C}}(\cdot,\cdot) - \inner{(\mathcal{C}+\alpha)^{\frac14}\cdot,  \widetilde{B}^\ast F \widetilde{B} (\mathcal{C}+\alpha)^{\frac14} \cdot},  \\
    &\sigma_1(H_\#^2) = \{ \lambda \in \R : \widetilde{B} (\mathcal{C}+\alpha)^{\frac14} P^{H_\#^2}(B_\epsilon(\lambda)) \neq 0 \quad \mbox{for small  $\epsilon>0$}\}, \quad\mbox{and}  \\
    &\sigma_1(\mathcal{C}) = \{\lambda \in \R: \widetilde{B} P^{\mathcal{C}}(B_\epsilon(\lambda)) \neq 0 \quad\mbox{for small $\epsilon>0$}\}, 
\end{align*}
where
\begin{align}
    \widetilde{B} \coloneqq B H_\#^{\frac12} (\mathcal{C}+\alpha)^{-\frac14}  \label{eq:Btildedef}
\end{align}
is bounded on $\{\Psi_0\}^\perp$ because $\mathcal{H}^{\frac12}(\mathcal{C}) = \mathcal{H}^1(H_\#)$. So repeating the same steps from before with the roles of $\mathcal{C}$ and $H_\#^2$ exchanged, the result follows.
\end{proof}

Next, we use the compactness assumption from Theorem~\ref{thm:maximaldomain} to show that the essential spectrum of $\mathcal{C}\rvert_{V_H}$,respectively, $\mathcal{C}\rvert_{V_\mathcal{C}}$ is equal to the essential spectrum of $H_\#^2 \rvert_{V_H}$, respectively, $H_\#^2\rvert_{V_\mathcal{C}}$.
\begin{lemma}[Invariance of essential spectrum] Suppose that $B^\ast F B \in \mathcal{B}_\infty(D(H_\#),\{\Psi_0\}^\perp)$, then
\begin{align}
    \sigma^{\mathrm{ess}}(\mathcal{C}\rvert_{V_H}) = \sigma^{\mathrm{ess}}(H_\#^2 \rvert_{V_H})\quad \mbox{and} \quad \sigma^{\mathrm{ess}}(H_\#^2\rvert_{V_\mathcal{C}}) = \sigma^{\mathrm{ess}}(\mathcal{C}\rvert_{V_{\mathcal{C}}}). \label{eq:reducedessentialspec}
\end{align}
\end{lemma}

\begin{proof} First, note that from Lemma~\ref{lem:CandC(z)} and the resolvent identity in Lemma~\ref{lem:resolventid} we have
\begin{align}
    (\mu + \mathcal{C})^{-1} = (\mu + H_\#^2)^{-1} + (\mu+\mathcal{C})^{-1} H_\#^{\frac12} B^\ast F B H_\#^{\frac12} (\mu + H_\#^2)^{-1} \label{eq:resolventequality}
\end{align}
as a bounded operator from $\mathcal{H}^{-1}(H_\#)$ to $\mathcal{H}^3(H_\#)$ for any $\mu >0$ big enough. In particular, by recalling the chain of inclusions in eq.~\eqref{eq:inclusionchain}, identity~\eqref{eq:resolventequality} holds in $\mathcal{B}(\{\Psi_0\}^\perp)$.  Furthermore, from Lemma~\ref{lem:reducingC} we have the block decomposition
\begin{align*}
    (\mu + \mathcal{C})^{-1} = \begin{pmatrix} (\mu + H_\#^2)^{-1}\rvert_{V_H^\perp} & 0 \\
    0 & (\mu+H_\#^2)^{-1}\rvert_{V_H} + (\mu+\mathcal{C})^{-1} H_\#^{\frac12} B^\ast F B H_\#^{\frac12} (\mu + H_\#^2)^{-1} \rvert_{V_H}\end{pmatrix}.
\end{align*}
Since $B^\ast F B \in \mathcal{B}_\infty(D(H_\#), \{\Psi_0\}^\perp)$, the second term in \eqref{eq:resolventequality} is compact in $\mathcal{B}(\{\Psi_0\}^\perp)$. Therefore, from Weyl's criterion, we conclude that
\begin{align*}
    \sigma^{\mathrm{ess}}((\mu+\mathcal{C})^{-1}\rvert_{V_H}) = \sigma^{\mathrm{ess}}((\mu+H_\#^2)^{-1}\rvert_{V_H}).
\end{align*}
Moreover, a similar argument shows that
\begin{align*}
    \sigma^{\mathrm{ess}}\bigr((\mu+\mathcal{C})^{-1}\rvert_{V_\mathcal{C}}\bigr) = \sigma^{\mathrm{ess}}\bigr((\mu+H_\#^2)^{-1}\rvert_{V_{\mathcal{C}}}\bigr).
\end{align*}
Eq.~\eqref{eq:reducedessentialspec} now follows from the relations $\sigma(f(A)) = f(\sigma(A))$ and $\ker f(\lambda) - f(A) = \ker \lambda - A$ (for injective $f$), which is a well-known corollary of the spectral theorem.\end{proof}

\begin{remark*}[Weaker compactness assumption] From Lemma~\ref{lem:CandC(z)} and the proof above, the weaker assumption $B^\ast F B \in \mathcal{B}_\infty\bigr(\mathcal{H}^3(H_\#),\mathcal{H}^{-2}(H_\#)\bigr)$ is actually enough to prove Theorem~\ref{thm:maximaldomain}.
\end{remark*}
We can now complete the proof of Theorem~\ref{thm:maximaldomain}.
\begin{proof}[Proof of Theorem~\ref{thm:maximaldomain}] Since $\sigma_1$ is closed, from the spectral theorem and the definitions of $V_\mathcal{C}$ and $V_H$ we have
\begin{align*}
    \sigma_1(\mathcal{C}) = \sigma(\mathcal{C}\rvert_{V_\mathcal{C}})\quad\mbox{and}\quad \sigma_1(H_\#^2) = \sigma(H_\#^2\rvert_{V_H}).
\end{align*}
In particular, we have
\begin{align*}
    \sigma_1^{\mathrm{ess}}(\mathcal{C}) = \sigma^{\mathrm{ess}}(\mathcal{C}\rvert_{V_\mathcal{C}}) \overset{\eqref{eq:reducedessentialspec}}{=}\sigma^{\mathrm{ess}}(H_\#^2\rvert_{V_\mathcal{C}}) \quad\mbox{and}\quad \sigma_1^{\mathrm{ess}}(H_\#^2) = \sigma^{\mathrm{ess}}(H_\#^2\rvert_{V_H}) \overset{\eqref{eq:reducedessentialspec}}{=} \sigma^{\mathrm{ess}}(\mathcal{C}\rvert_{V_H}).
\end{align*}
Hence to conclude the proof, it is enough to show that
\begin{align}
    \sigma_1^{\mathrm{ess}}(\mathcal{C}) \subset \sigma^{\mathrm{ess}}(\mathcal{C}\rvert_{V_H}) \quad\mbox{and}\quad \sigma^{\mathrm{ess}}_1(H_\#^2) \subset \sigma^{\mathrm{ess}}(H_\#^2\rvert_{V_\mathcal{C}}). \label{eq:midinclusion}
\end{align}
So suppose that $\lambda \not \in \sigma^{\mathrm{ess}}(\mathcal{C}\rvert_{V_H})$. Then $\lambda$ can be, at most, an isolated point in the spectrum of $\mathcal{C}\rvert_{V_H}$, which implies that
\begin{align*}
    P^{\mathcal{C}}(B_\epsilon(\lambda)) = P^{\mathcal{C}\rvert_{V_H}}(\{\lambda\}) + P^{\mathcal{C}\rvert_{V_H^\perp}}(B_\epsilon(\lambda)) \quad\mbox{( for $\epsilon>0$ small).}
\end{align*}
From the fact that $P_{V_H^\perp}$ commutes with $H_\#$ and $BP_{V_H^\perp} = 0$ (see eq.~\eqref{eq:BVH}), we find that
\begin{align*}
     \rank B H_\#^{\frac12} P^{\mathcal{C}}(B_\epsilon(\lambda)) =  \rank B H_\#^{\frac12} P^{\mathcal{C}\rvert_{V_H}}(\{\lambda\}) \leq \rank P^{\mathcal{C}\rvert_{V_H}}(\{\lambda\}) < \infty,
\end{align*}
which shows that $\lambda \not \in \sigma_1^{\mathrm{ess}}(\mathcal{C})$. The second inclusion in \eqref{eq:midinclusion} follows from the same arguments with the roles of $H_\#^2$ and $\mathcal{C}$ interchanged. \end{proof}


\subsection{Proof of Proposition~\ref{prop:compactnesscriterion}}

For the proof of Proposition~\ref{prop:compactnesscriterion}, we shall use the following version of the Rellich-Kondrachov (aka compact Sobolev embedding) theorem.
\begin{lemma}[Rellich-Kondrachov Theorem]\label{lem:RKthm} Let $\{\Psi_j\}_{j \in \N} \subset \mathcal{H}^1(\Delta) = \{ \Psi: \R^n \rightarrow \C : \norm{\Psi}_{L^2(\R^n)} + \norm{\nabla \Psi}_{L^2(\R^n)} < \infty\}$ be a bounded sequence. Then we can extract a subsequence such that for any $\Omega \subset \R^n$ with finite (Lebesgue) measure, we have
\begin{align}
    \int_{\Omega} |\Phi_j(r) - \Phi_k(r)|^2 \mathrm{d} r \ra 0 \quad \mbox{as } \min\{j,k\} \ra \infty . \label{eq:Rellich}
\end{align}
\end{lemma}
\begin{proof} By the standard compact Sobolev embedding (see \cite[Theorem 1, Section 5.7]{EvansPDE}) and a standard diagonal argument, we can extract a subsequence such that \eqref{eq:Rellich} holds for $\Omega = B_M$ for any radius $M>0$. Thus from H\"older's inequality,
\begin{align*}
    \int_{\Omega}|\Phi_j - \Phi_k|^2 \mathrm{d}r \lesssim \int_{\Omega \cap B_M} |\Phi_j-\Phi_k|^2 \mathrm{d}r + |\Omega\setminus B_M|^{1-\frac{2}{p}} \norm{\Phi_j - \Phi_k}_{L^p}^{\frac{2}{p}} \quad \mbox{for $p \in [2,\infty]$.}
\end{align*}
From the classical Sobolev embedding, the norms $\norm{\Phi_j-\Phi_k}_{L^p}$ are uniformly bounded (in $j,k$) for any $2 < p \leq \frac{2n}{n-2}$. So choosing $M>0$ arbitrarily large, the second term can be made arbitrarily small, which completes the proof.
\end{proof}

\begin{proof}[Proof of Proposition~\ref{prop:compactnesscriterion}]
For the RPA, we have $B^\ast F^{\mathrm{RPA}} B = P_{\Psi_0^\perp} T$, where $T$ is an integral operator with integral kernel given by
\begin{align*}
    T(r_1,..,r_N, r_1',...,r_N') =  N\sum_{j=1}^N \frac{\overline{\Psi_0(r_1,...,r_N)}\Psi_0(r_1',...,r_N')}{|r_j - r_1'|}.
\end{align*}
Then from Cauchy-Schwarz,
\begin{align*}
    \norm{T}_{L^2(\R^{3N}\times \R^{3N})}^2 \lesssim \int_{\R^6} \frac{\rho_0(r) \rho_0(r')}{|r-r'|^2} \mathrm{d}r\mathrm{d}r' \lesssim \norm{\rho_0}_{L^1\cap L^\infty}^2 \norm{|\cdot|^{-2}}_{L^1+L^\infty} < \infty,
\end{align*}
which implies that $T$ is Hilbert-Schmidt, and therefore compact (even in $\mathcal{B}_\infty(\{\Psi_0\}^\perp$).

For the operator $F_{\rho_0}$, we first note that $B^\ast F_{\rho_0} B = B^\ast F_{\rho_0} SP_{\Psi_0^\perp}$ where
\begin{align*}
    S \Phi(r) = N\int_{\R^{3N-3}} \overline{\Psi_0(r,r_2,...,r_N)} \Phi(r, r_2,...,r_N) \mathrm{d}r_2 ... \mathrm{d} r_N.
\end{align*}
In particular, from the assumption that $D(H_\#)$ is continuously embedded in $\mathcal{H}^1(\Delta)$, it is enough to show that for any bounded sequence $\{\Phi_j \}_{j \in \N}$ in $ \mathcal{H}^1(\Delta)$, there exists a subsequence that satisfies (after re-labeling the indices)
\begin{align}
    \norm{F_{\rho_0} S (\Phi_j-\Phi_k)}_{L^2_{\rho_0}}^2 \ra 0 \quad\mbox{as $\min\{j,k\} \ra \infty$.}\label{eq:subsequence}
\end{align}
 To find such a subsequence, we apply the Rellich-Kondrachov theorem (Lemma~\ref{lem:RKthm}). Precisely, let $\{\Phi_j\}$ be the subsequence satisfying \eqref{eq:Rellich} and define 
\begin{align*}
    I_\epsilon \coloneqq \{ r \in \R^3 : \rho_0(r) > \epsilon \}.
\end{align*}
Then by the Cauchy-Schwarz inequality (as in \eqref{eq:CSineq}) and assumption~\eqref{eq:ALDAassump} we obtain
\begin{align*}
\norm{F_{\rho_0}S(\Phi_j- \Phi_j)}_{L^2_{\rho_0}}^2 &\lesssim \int_{\R^3} \rho_0(r)^{2\delta+1} \biggr|\int_{\R^{3N-3}} \overline{\Psi_0(r,\tilde{r})} \bigr(\Phi_j(r,\tilde{r})-\Phi_k(r,\tilde{r})\bigr) \mathrm{d} \tilde{r}\biggr|^2 \mathrm{d}r \\
&\lesssim \int_{\R^3 \setminus I_\epsilon} \rho_0(r)^{2\delta+2} \rho_{\Phi_j-\Phi_k}(r) \mathrm{d}r + \norm{\rho_0}_{L^\infty}^{2\delta+2}\int_{I_\epsilon \times B_M} 
 \bigr|\Phi_j(r,\tilde{r})-\Phi_k(r,\tilde{r})\bigr|^2\mathrm{d}r\mathrm{d}\tilde{r} \\
 &+ \int_{I_\epsilon} \rho_0(r)^{2\delta+1}\biggr|\int_{\R^{3N-3}\setminus B_M} \Psi_0(r,\tilde{r}) \bigr(\Phi_j - \Phi_k\bigr)(r,\tilde{r}) \mathrm{d}\tilde{r} \biggr|^2 \mathrm{d} r,
\end{align*}
where $B_M$ is the ball of radius $M$ centered at the origin. From the definition of $I_\epsilon$, the first term is bounded by $\lesssim \epsilon^{2\delta+2}$. Moreover, the $\R^3$-measure of $I_\epsilon$ is finite (because $\rho_0 \in L^1$) and the second term vanishes as $\min\{j,k\} \ra \infty$ by \eqref{eq:Rellich}. Lastly, we can bound the third term by 
\begin{multline*}
    \int_{I_\epsilon} \rho_0(r)^{2\delta +1} \biggr(\int_{\R^{3N-3}\setminus B_M} \Psi_0(r,\tilde{r})\bigr(\Phi_j - \Phi_k)(r,\tilde{r})\mathrm{d} \tilde{r}\biggr)^2 \mathrm{d}r \\
    \leq \max \{\epsilon^{-2\delta -1}, \norm{\rho_0}_{L^\infty}^{2\delta+1}\} \int_{I_\epsilon} \rho_{0,M}(r) \rho_{\Phi_j-\Phi_k}(r) \mathrm{d}r \leq C(\epsilon, \rho_0) \norm{\rho_{0,M}}_{L^p} \norm{\rho_{\Phi_j-\Phi_k}}_{L^q}
\end{multline*}
for any $p^{-1} + q^{-1} = 1$, where 
\begin{align*}
    \rho_{0,M}(r) = \int_{\R^{3N-3}\setminus B_M} |\Psi_0(r,\tilde{r})|^2 \mathrm{d}\tilde{r}.
\end{align*}
Therefore, from the inequality
\begin{align*}
    \norm{\nabla \sqrt{\rho_{\Phi}}}_{L^2(\R^3)} \lesssim \norm{\nabla \Phi}_{L^2(\R^{3N})}
    \quad \mbox{(see \cite{lieb1983} for a proof)}
\end{align*}
and the classical Sobolev embedding in $\R^3$, the norms $\norm{\rho_{\Phi_j-\Phi_k}}_{L^q(\R^3)}$ are uniformly bounded for any $1 \leq q \leq 3$. Consequently, by dominated convergence (recall that $\rho \in L^1 \cap L^\infty$), the last term can be made arbitrarily small by choosing $M$ large.
\end{proof}

\appendix

\section{Optimality of weighted density spaces} \label{sec:weighteddensity}

We have shown that $\chi_H(t) \in \mathcal{B}(E^\star ,E)$ for the spaces $E = L^2_{\nicefrac{1}{\rho_0}}$ and $E = L^1(\Omega, \mathrm{d}\mu)$ (see eq.~\eqref{eq:CSineq}). In addition, if $\rho_0 \in L^1 \cap L^\infty$, then we can also choose $E = L^1(\Omega,\mathrm{d}\mu) \cap L^2(\Omega,\mathrm{d}\mu)$ by H\"older's inequality. Moreover, in this case we have the inclusions
\begin{align*}
    L^2_{\nicefrac{1}{\rho_0}} \subset L^1(\Omega,\mathrm{d}\mu) \cap L^2(\Omega,\mathrm{d}\mu) \subset L^1(\Omega, \mathrm{d} \mu) .
\end{align*}
Hence a natural question is whether $E = L^2_{\nicefrac{1}{\rho_0}}$ is a minimal space for which $\chi_H \in C_s\bigr(\R, \mathcal{B}(E^\star,E)\bigr)$. This question is not only natural but also relevant because a minimal $E$ yields a maximal space of allowed adiabatic approximations $\mathcal{B}(E,E^\star)$. 

We now give a partial answer to this question. The idea is the following. Looking back at the definition of $\chi_H$, we see from a duality argument that $\chi_H(t) \in \mathcal{B}(E^\star,E)$ as long as we can show that $B : \{\Psi_0\}^\perp \rightarrow E$ is bounded. Thus a reasonable approach is to look for a minimal subspace $E$ for which $B:\{\Psi_0\}^\perp \rightarrow E$ is bounded. It turns out that $E = L^2_{\nicefrac{1}{\rho_0}}$ is minimal among a general class of function spaces.

\begin{proposition}[Minimality of $L^2_{\nicefrac{1}{\rho_0}}$]\label{prop:maximalL^2rho0} Let $E$ be a Banach space of $\mu$-measurable functions such that \begin{enumerate}[label=(\roman*)]
\item $\rho_0 \in E$ and \item $E$ has the lattice property , i.e., for any measurable $g$ with $|g| \leq |f|$ a.e. for some $f \in E$, we have $g \in E$ and $\norm{g}_E \leq \norm{f}_E$. 
\end{enumerate}
Then if $B : \{\Psi_0\}^\perp \rightarrow E$ is bounded and $E \subset L^2_{\nicefrac{1}{\rho_0}}$, we have $E = L^2_{\nicefrac{1}{\rho_0}}$.
\end{proposition}
\begin{proof} Let $f \in L^2_{\nicefrac{1}{\rho_0}}$ and define $\Phi_f \coloneqq B^\ast |f| \in \{\Psi_0\}^\perp$. Then the function
\begin{align*}
    B \Phi_f = |f(r)| + N(N-1)\int_{\Omega^{N-1}} |f(r_2)| \frac{|\Psi_0(r,r_2,...,r_N)|^2}{\rho_0(r)}\mathrm{d}\mu(r_2)...\mathrm{d}\mu(r_N) - N^2 \inner{1,f} \rho_0(r) 
\end{align*}
belongs to $E$ by assumption. But since $\rho_0 \in E$, we have $|f| \leq B \Phi_f + N^2\inner{1,f} \rho_0 \in E$, which implies that $f \in E$ by the lattice property.
\end{proof}

In particular, we see that the space $L^2_{\nicefrac{1}{\rho_0}}$ is minimal over the large class of Banach function spaces \cite[Chapter 6]{FunctionSpaces2013}.

\begin{remark*}[Reduced weighted density spaces] The weighted density space $L^2_{\nicefrac{1}{\rho_0}}$ is in fact the range of the operator $B$ when extended to the whole tensor product space $\otimes_{j=1}^N L^2(\Omega,d\mu)$ via eq.~\eqref{eq:Bdef}. When restricting this extension to the orthogonal complement $\{\Psi_0\}^\perp$ on $\otimes_{j=1}^N L^2(\Omega,d\mu)$, the range of $B$ is given by the annihilator of $1 \in L^2_{\rho_0}$, i.e., 
\begin{align*}
    1^\perp \coloneqq \biggr\{ f \in L^2_{\nicefrac{1}{\rho_0}} : \int_\Omega f(r) \mathrm{d} \mu(r) = 0\biggr\}.
\end{align*}
In particular, we could replace the spaces $L^2_{\nicefrac{1}{\rho_0}}$ and $L^2_{\rho_0}$, respectively, by $1^\perp$ and the quotient space
\begin{align*}
    (1^\perp)^\star = L^2_{\rho_0}/1 = \{ [f] : f \sim g \quad \mbox{if and only if}\quad  f(r) - g(r) = \mbox{constant $\mu$-a.e.}\}
\end{align*}
with the induced norm. This choice of spaces incorporates the fact that possible variations of the density have zero average, thus preserving the number of particles in the system, and the fact that potentials differing by a constant give the same variation of the density. The main reason for working with the spaces $L^2_{\nicefrac{1}{\rho_0}}$ and $L^2_{\rho_0}$ instead is that they simplify the presentation.
\end{remark*}

\bibliographystyle{abbrv}
\bibliography{Extendedreferences}

\begin{thebibliography}{10}

\bibitem{agmon1975spectral}
S.~Agmon.
\newblock Spectral properties of {S}chr\"{o}dinger operators and scattering
  theory.
\newblock {\em Ann. Scuola Norm. Sup. Pisa Cl. Sci. (4)}, 2(2):151--218, 1975.

\bibitem{agmon2014lectures}
S.~Agmon.
\newblock {\em Lectures on exponential decay of solutions of second-order
  elliptic equations: bounds on eigenfunctions of {$N$}-body {S}chr\"{o}dinger
  operators}, volume~29 of {\em Mathematical Notes}.
\newblock Princeton University Press, Princeton, NJ; University of Tokyo Press,
  Tokyo, 1982.

\bibitem{commutatormethodsbook}
W.~O. Amrein, A.~Boutet~de Monvel, and V.~Georgescu.
\newblock {\em {$C_0$}-groups, commutator methods and spectral theory of
  {$N$}-body {H}amiltonians}, volume 135 of {\em Progress in Mathematics}.
\newblock Birkh\"{a}user Verlag, Basel, 1996.

\bibitem{BaiCangLReigenvalueI2012}
Z.~Bai and R.-C. Li.
\newblock Minimization principles for the linear response eigenvalue problem
  {I}: {T}heory.
\newblock {\em SIAM J. Matrix Anal. Appl.}, 33(4):1075--1100, 2012.

\bibitem{BaiCangLReigenvalueII2013}
Z.~Bai and R.-C. Li.
\newblock Minimization principles for the linear response eigenvalue problem
  {II}: {C}omputation.
\newblock {\em SIAM J. Matrix Anal. Appl.}, 34(2):392--416, 2013.

\bibitem{Brabec2015}
J.~Brabec, L.~Lin, M.~Shao, N.~Govind, C.~Yang, Y.~Saad, and E.~G. Ng.
\newblock Efficient algorithms for estimating the absorption spectrum within
  linear response tddft.
\newblock {\em Journal of Chemical Theory and Computation}, 11(11):5197--5208,
  Nov 2015.

\bibitem{burkeDFTreview}
K.~Burke, J.~Werschnik, and E.~Gross.
\newblock Time-dependent density functional theory: Past, present, and future.
\newblock {\em The Journal of chemical physics}, 123:62206, 09 2005.

\bibitem{GW0method2016}
E.~Canc\'{e}s, D.~Gontier, and G.~Stoltz.
\newblock A mathematical analysis of the {$\rm{GW}^0$} method for computing
  electronic excited energies of molecules.
\newblock {\em Rev. Math. Phys.}, 28(4):1650008, 51, 2016.

\bibitem{CancesHartree1999}
E.~Canc\`es and C.~Le~Bris.
\newblock On the time-dependent {H}artree-{F}ock equations coupled with a
  classical nuclear dynamics.
\newblock {\em Math. Models Methods Appl. Sci.}, 9(7):963--990, 1999.

\bibitem{CancesRPA2012}
E.~Canc\`es and G.~Stoltz.
\newblock A mathematical formulation of the random phase approximation for
  crystals.
\newblock {\em Ann. Inst. H. Poincar\'{e} C Anal. Non Lin\'{e}aire},
  29(6):887--925, 2012.

\bibitem{CASIDA2009}
M.~E. Casida.
\newblock Time-dependent density-functional theory for molecules and molecular
  solids.
\newblock {\em Journal of Molecular Structure: THEOCHEM}, 914(1):3--18, 2009.
\newblock Time-dependent density-functional theory for molecules and molecular
  solids.

\bibitem{CeperleyQMCcorrelation}
D.~M. Ceperley and B.~J. Alder.
\newblock Ground state of the electron gas by a stochastic method.
\newblock {\em Phys. Rev. Lett.}, 45:566--569, Aug 1980.

\bibitem{ChadamHartree1975}
J.~M. Chadam and R.~T. Glassey.
\newblock Global existence of solutions to the {C}auchy problem for
  time-dependent {H}artree equations.
\newblock {\em J. Mathematical Phys.}, 16:1122--1130, 1975.

\bibitem{DDRF2023}
T.~C. Corso, M.-S. Dupuy, and G.~Friesecke.
\newblock The density-density response function in time-dependent density
  functional theory: mathematical foundations and pole shifting.
\newblock {\em arXiv preprint arXiv:2301.13070}, 2023.

\bibitem{dupuy:hal-03145143}
M.-S. Dupuy and A.~Levitt.
\newblock Finite-size effects in response functions of molecular systems.
\newblock {\em SMAI J. Comput. Math.}, 8:273--294, 2022.

\bibitem{dyatlov2019mathematical}
S.~Dyatlov and M.~Zworski.
\newblock {\em Mathematical theory of scattering resonances}, volume 200 of
  {\em Graduate Studies in Mathematics}.
\newblock American Mathematical Society, Providence, RI, 2019.

\bibitem{EvansPDE}
L.~C. Evans.
\newblock {\em Partial differential equations}, volume~19 of {\em Graduate
  Studies in Mathematics}.
\newblock American Mathematical Society, Providence, RI, second edition, 2010.

\bibitem{lewin2014hartree}
M.~Lewin and J.~Sabin.
\newblock The {H}artree equation for infinitely many particles, {II}:
  Dispersion and scattering in 2d.
\newblock {\em Analysis \& PDE}, 7(6):1339--1363, 2014.

\bibitem{lieb1983}
E.~H. Lieb.
\newblock Density functionals for {C}oulomb systems.
\newblock {\em International Journal of Quantum Chemistry}, 24(3):243--277,
  1983.

\bibitem{lin2019mathematical}
L.~Lin and J.~Lu.
\newblock {\em A mathematical introduction to electronic structure theory},
  volume~4 of {\em SIAM Spotlights}.
\newblock Society for Industrial and Applied Mathematics (SIAM), Philadelphia,
  PA, 2019.

\bibitem{marques2012fundamentals}
M.~A.~L. Marques, N.~T. Maitra, F.~M.~S. Nogueira, E.~K.~U. Gross, and
  A.~Rubio, editors.
\newblock {\em Fundamentals of time-dependent density functional theory},
  volume 837 of {\em Lecture Notes in Physics}.
\newblock Springer, Heidelberg, 2012.

\bibitem{OLSEN1988}
J.~Olsen, H.~J.~A. Jensen, and P.~Jørgensen.
\newblock Solution of the large matrix equations which occur in response
  theory.
\newblock {\em Journal of Computational Physics}, 74(2):265--282, 1988.

\bibitem{PW92}
J.~P. Perdew and Y.~Wang.
\newblock Accurate and simple analytic representation of the electron-gas
  correlation energy.
\newblock {\em Phys. Rev. B}, 45:13244--13249, Jun 1992.

\bibitem{PGG1996}
M.~Petersilka, U.~J. Gossmann, and E.~K.~U. Gross.
\newblock Excitation energies from time-dependent density-functional theory.
\newblock {\em Phys. Rev. Lett.}, 76:1212--1215, Feb 1996.

\bibitem{FunctionSpaces2013}
L.~Pick, A.~Kufner, O.~John, and S.~Fu\v{c}\'{\i}k.
\newblock {\em Function spaces. {V}ol. 1}, volume~14 of {\em De Gruyter Series
  in Nonlinear Analysis and Applications}.
\newblock Walter de Gruyter \& Co., Berlin, extended edition, 2013.

\bibitem{pusateri2021long}
F.~Pusateri and I.~M. Sigal.
\newblock Long-time behaviour of time-dependent density functional theory.
\newblock {\em Archive for Rational Mechanics and Analysis}, 241(1):447--473,
  2021.

\bibitem{ReedSimonI}
M.~Reed and B.~Simon.
\newblock {\em Methods of modern mathematical physics. {I}. {F}unctional
  analysis}.
\newblock Academic Press, New York-London, 1972.

\bibitem{ReedSimonII}
M.~Reed and B.~Simon.
\newblock {\em Methods of modern mathematical physics. {II}. {F}ourier
  analysis, self-adjointness}.
\newblock Academic Press [Harcourt Brace Jovanovich, Publishers], New
  York-London, 1975.

\bibitem{ReedSimonIV}
M.~Reed and B.~Simon.
\newblock {\em Methods of modern mathematical physics. {IV}. {A}nalysis of
  operators}.
\newblock Academic Press [Harcourt Brace Jovanovich, Publishers], New
  York-London, 1978.

\bibitem{BookSchmudgen2012}
K.~Schm\"{u}dgen.
\newblock {\em Unbounded self-adjoint operators on {H}ilbert space}, volume 265
  of {\em Graduate Texts in Mathematics}.
\newblock Springer, Dordrecht, 2012.

\bibitem{SimonSchrSemigroup82}
B.~Simon.
\newblock Schr\"{o}dinger semigroups.
\newblock {\em Bull. Amer. Math. Soc. (N.S.)}, 7(3):447--526, 1982.

\bibitem{SprengelTDKS2017}
M.~Sprengel, G.~Ciaramella, and A.~Borz\`i.
\newblock A theoretical investigation of time-dependent {K}ohn-{S}ham
  equations.
\newblock {\em SIAM J. Math. Anal.}, 49(3):1681--1704, 2017.

\bibitem{THOULESS1961}
D.~J. Thouless.
\newblock Vibrational states of nuclei in the random phase approximation.
\newblock {\em Nuclear Phys.}, 22:78--95, 1961.

\bibitem{Toulouse2022}
J.~Toulouse, K.~Schwinn, F.~Zapata, A.~Levitt, E.~Canc\'{e}s, and E.~Luppi.
\newblock {Photoionization and core resonances from range-separated
  time-dependent density-functional theory for open-shell states: Example of
  the lithium atom}.
\newblock {\em The Journal of Chemical Physics}, 157(24), 12 2022.
\newblock 244104.

\bibitem{ullrich2012time}
C.~Ullrich.
\newblock {\em Time-Dependent Density-Functional Theory: Concepts and
  Applications}.
\newblock Oxford Graduate Texts. OUP Oxford, 2012.

\bibitem{Vasiliev2002}
I.~Vasiliev, S.~\"O\ifmmode~\breve{g}\else \u{g}\fi{}\"ut, and J.~R.
  Chelikowsky.
\newblock First-principles density-functional calculations for optical spectra
  of clusters and nanocrystals.
\newblock {\em Phys. Rev. B}, 65:115416, Mar 2002.

\bibitem{ZangwillSoven80}
A.~Zangwill and P.~Soven.
\newblock Density-functional approach to local-field effects in finite systems:
  Photoabsorption in the rare gases.
\newblock {\em Phys. Rev. A}, 21:1561--1572, May 1980.

\end{thebibliography}

\end{document}